\documentclass[conference]{IEEEtran}
\IEEEoverridecommandlockouts
\usepackage{cite}
\usepackage{amsmath,amssymb,amsfonts}
\usepackage{textcomp}

\usepackage{cite}
\usepackage{amsmath,amssymb,amsfonts}
\usepackage{graphicx}
\usepackage{textcomp}
\usepackage{multirow}
\usepackage{subcaption}
\usepackage{hyperref}  
\usepackage{xcolor}
\usepackage{algorithm}
\usepackage{algpseudocode}
\usepackage{pifont}
\usepackage[normalem]{ulem}
\usepackage{listings}
\usepackage{xcolor}

\newcommand{\etal}{\emph{et al.\ }}

\newcommand{\fixit}[1]{\textcolor{black}{#1}}

\newcommand{\eddy}[1]{\textcolor{black}{#1}}

\def\BibTeX{{\rm B\kern-.05em{\sc i\kern-.025em b}\kern-.08em
    T\kern-.1667em\lower.7ex\hbox{E}\kern-.125emX}}

\begin{document}

\title{Tetris: A Compilation Framework for VQA
Applications in Quantum Computing\\

\thanks{* These two authors have equal contributions. }
}

\author{

\IEEEauthorblockN{ Yuwei Jin*}
\IEEEauthorblockA{
\textit{yj243@cs.rutgers.edu}\\
\textit{Rutgers University}\\
USA }
\and
\IEEEauthorblockN{ Zirui Li*}
\IEEEauthorblockA{
\textit{zirui.li@rutgers.edu}\\
\textit{Rutgers University}\\
USA }
\and
\IEEEauthorblockN{Fei Hua}
\IEEEauthorblockA{
\textit{huafei90@gmail.com}\\
\textit{Rutgers University}\\
USA }
\and
\IEEEauthorblockN{Tianyi Hao}
\IEEEauthorblockA{
\textit{tianyi.hao@wisc.edu}\\
\textit{University of Wisconsin-Madison}\\
USA }
\and

\IEEEauthorblockN{Huiyang Zhou}
\IEEEauthorblockA{
\textit{hzhou@ncsu.edu}\\
\textit{North Carolina State University}\\
USA }
\and
\IEEEauthorblockN{Yipeng Huang}
\IEEEauthorblockA{
\textit{yipeng.huang@rutgers.edu}\\
\textit{Rutgers University}\\
USA }
\and
\IEEEauthorblockN{Eddy Z. Zhang}
\IEEEauthorblockA{\textit{eddy.zhengzhang@gmail.com} \\
\textit{Rutgers University}\\
USA }

}

\maketitle

\begin{abstract}
Quantum computing has shown promise in solving complex problems by leveraging the principles of superposition and entanglement. Variational quantum algorithms (VQA) are a class of algorithms suited for near-term quantum computers due to their modest requirements of qubits and depths of computation. This paper introduces Tetris -- a compilation framework for VQA applications on near-term quantum devices. Tetris focuses on reducing two-qubit gates in the compilation process since a two-qubit gate has an order of magnitude more significant error and execution time than a single-qubit gate. Tetris exploits unique opportunities in the circuit synthesis stage often overlooked by the state-of-the-art VQA compilers for reducing the number of two-qubit gates. Tetris comes with a refined IR of Pauli string to express such a two-qubit gate optimization opportunity.
Moreover, Tetris is equipped with a fast bridging approach that mitigates the hardware mapping cost. Overall, Tetris demonstrates a reduction of {up to 41.3\%} in CNOT gate counts,  37.9\% in circuit depth, and { 42.6\%} in circuit duration for various molecules of different sizes and structures compared with the state-of-the-art approaches. Tetris is open-sourced at this 
\href{https://github.com/abclzr/vqe_tetris}{link}. 
\end{abstract}


\section{Introduction}

Quantum computing, a rapidly evolving field, has the potential to revolutionize the way we process information by leveraging the principles of quantum mechanics. Among the various applications of quantum computing, variational quantum applications  \cite{raeisi+:physicasjournal12, poulin_trotter:arxiv14,jordan+:science12,hempel+:PhysRevX18, arute+:arxiv2020,georgescu+:revmodphys2014} have emerged as a promising area for achieving near-term quantum advantage. The idea is to provide a hybrid quantum-classical framework with a parameterized quantum ansatz circuit and a classical optimizer that iteratively tunes the parameters of the ansatz circuit. Examples include Variational Quantum Eigensolver (VQE) \cite{peruzzo+:nature14} and the Quantum Approximate Optimization Algorithm (QAOA) \cite{farhi+:arxiv16,farhi+2014quantum,farhi+:quantum22, lao+:isca22,alam+:micro20,jin+:asplos23_qaoa}. 

\vspace{0.05in}
\noindent \textbf{Compilation for Variational Quantum Algorithms:} Variational quantum algorithms (VQA) typically have ansatz circuits that are generated from a matrix exponential $e^{-i\cdot c \cdot H}$, where $H$ is the Hamiltonian, represented as a matrix, and $c$ is some constant as a product of scalars including the Planck constant in Schr\"odinger's equation. A matrix exponential in this form is a derived solution of a Schr\"odinger equation for the time evolution of a quantum system of interest. 

\begin{figure}[htb]
    \centering
\includegraphics[width=0.95\linewidth]{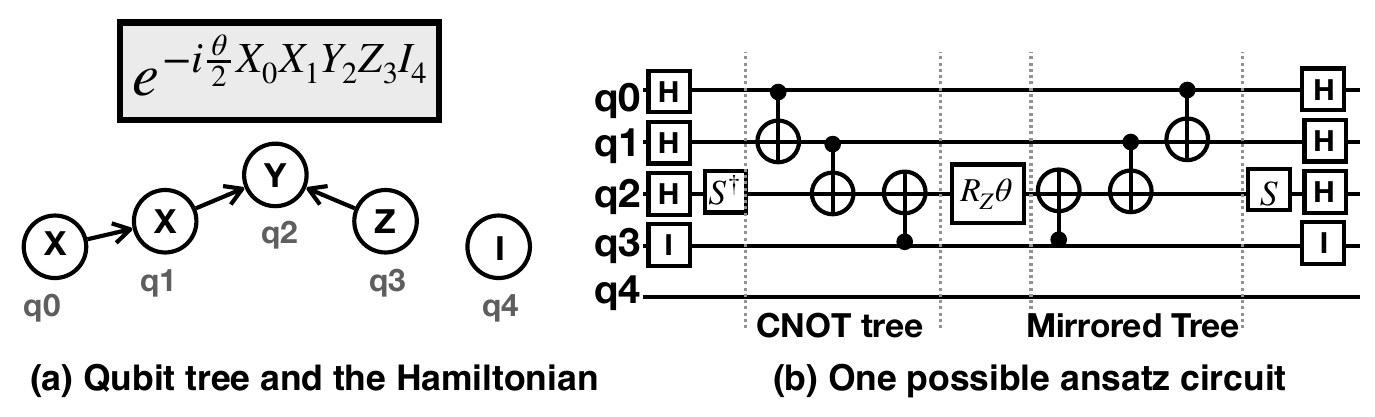}
    \caption{ For the  Pauli string $X_{0}X_{1}Y_{2}Z_{3}I_{4}$ and the corresponding matrix exponential $e^{-i\frac{\theta}{2} X_{0}X_{1}Y_{2}Z_{3}I_{4}}$, a qubit tree, and an ansatz circuit can be synthesized.}
    \label{fig:twovqe-egs}
\end{figure}

It is necessary to transform such a matrix exponential into a quantum circuit that can run on a real quantum computer. The transformation follows multiple steps \cite{peruzzo+:nature14} including the trotterization steps \cite{trotter:1959,suzuki:1991} and the Hamiltonian decomposition steps \cite{li+:asplos22, li+:isca21}. A well-known abstraction is to rewrite the Hamiltonian $H$ into a sum of simpler elements consisting of only tensor products of Pauli-operators, where Pauli-operators include $X$, $Y$, $Z$, and the identity operator $I$. The tensor product of Pauli-operators is typically represented as a Pauli-string, as shown in Fig. \ref{fig:twovqe-egs} (a). Each Pauli-string corresponds to a synthesized sub-circuit. Composing these sub-circuits together forms the ansatz circuit in a VQA application.

\vspace{0.05in}
\noindent \textbf{Tree-based Synthesis Rules:} We show an example of synthesizing a circuit to implement the matrix exponential $e^{-i\cdot\frac{\theta}{2}\cdot XXYZI}$ in Fig. \ref{fig:twovqe-egs} for the UCCSD ansatz \cite{peruzzo+:nature14} in computational chemistry. The scalar parameter $\theta$  appears as one-half of the rotation angle of the single-qubit rotation gate applied to the root qubit, which will be shown later. 

A Pauli-string ``XXYZI" has positional correspondence  to qubits $q_0$, $q_1$, $q_2$, $q_3$, and $q_4$ respectively. The qubits that have the identity I operators do not perform any gate. The rule is to construct a directed tree for the qubits corresponding to X, Y, or Z. Any qubit can be the root. The tree is valid as long as every other qubit can follow a directed path to the root qubit. A valid tree is shown in Fig. \ref{fig:twovqe-egs} (a).

From a tree, a valid circuit is constructed by scheduling the CNOT gates corresponding to directed edges in the tree, applying a single-qubit rotation gate to the root qubit, and mirroring CNOT gates after the single-qubit gate. If two edges (CNOTs) form a directed path, it creates a dependence between the two CNOTs. The direction indicates the dependence order. CNOTs can run in arbitrary order if they do not have any dependence. Two layers of single-qubit gates wrap around the CNOT tree sub-circuit generated above (also mirrored). The generation of single-qubit gates also follows specific rules with respect to a Pauli-string. They are described in prior work \cite{li+:isca21, li+:asplos22}. We will not describe them here as they are irrelevant to this paper. Fig. \ref{fig:twovqe-egs} (b) shows a complete generated circuit from the Pauli-string XXYZI.

\vspace{0.05in}
\noindent \textbf{Prior Work} A vast body of research has been conducted to reduce the size of the ansatz for variational quantum eigensolvers (VQE) \cite{tilly+:physicsreports22}, a lot of these studies \cite{lee+:jctc19, grimsley+:naturecomm19, dallairedemers+:vqearxiv18, ryabinkin+:vqe18arxiv, ryabinkin+:jctc20, tang+:vqe21arxiv} are theoretical, oblivious to the underlying hardware. Li \etal \cite{li+:isca21} \cite{li+:asplos22} propose the first compiler studies that consider co-optimization of architecture and software for variational algorithms in chemistry simulation and also for efficient hardware mapping to near-term superconducting devices with consideration of SWAP insertion, etc. 

However, in general, compiling VQA computation kernels is challenging. There is typically a large number of Pauli-strings in the ansatz circuit. For instance, the UCCSD ansatz uses $O(n^4)$ Pauli-strings, where $n$ corresponds to the number of qubits. It implies $O(n^4)$ sub-circuits need to be generated. Li \etal \cite{li+:asplos22, li+:isca21} focus on reducing the number of CNOT gates in the final compiled hardware-compliant circuits. A two-qubit gate in superconducting hardware has an order of magnitude higher latency and error than a single-qubit gate. Our paper focuses on the same goal of reducing the CNOT gate count.

\vspace{0.05in}
\noindent \textbf{Key Takeway Messages} 
We propose Tetris -- a compilation and optimization framework for variational quantum algorithms. Tetris is the first efficient compiler framework that \emph{systematically explores two-qubit gate optimization opportunities and its synergy with SWAP insertion}. We argue that two-qubit gate reduction not only results from reduced SWAP gates during hardware-aware compilation but also from the tree-based circuit synthesis rules during the Pauli-string synthesis stage. Compared with prior studies \cite{li+:asplos22, lao+:isca22, li+:isca21}, our key contributions are the following:

\ding{172} We discovered that prior work tends to prioritize SWAP insertion when synthesizing circuits to deal with hardware connectivity constraints.
While hardware-aware circuit synthesis is important, we should not overlook \emph{two-qubit-gate-canceling} aware synthesis. Mainly focusing on SWAP reduction might result in a significant amount of missed opportunities in canceling gates across Pauli-strings with high similarity.

\ding{173} Similarity across Pauli-strings not only results in 1-qubit gate cancellation, but also 2-qubit gate cancellation. We analyzed a number of representative VQE benchmarks for finding the minimal ground state energy of molecules using up to 20 thousand Pauli-strings and found out there is a significant opportunity for reducing 2-qubit gates, but such opportunity is often overlooked by the prior studies \cite{li+:asplos22, li+:isca21}. 

\ding{174}       
    While \emph{Tetris} improves the compiler's capability of 2-qubit gate canceling, it also carefully considers SWAP insertion and 1-qubit gate cancellation and strikes a balance between different optimization factors.

 \ding{175} We propose the Tetris-IR,  a refined IR of the Pauli-string representation. It provides a further abstraction of the VQA kernel for expressing the 2-qubit gate-canceling opportunity and facilitates later optimization when lowering the IR in the compiler stack. Additionally, we propose a \emph{fast bridging} approach to mitigate SWAP insertion cost when there are available mid-circuit measurement opportunities\cite{hua+:asplos2023}.  

    Overall, Tetris
demonstrates up to 41.3\% reduction in 2-qubit
gate counts, 37.9\% reduction in depth, 1.74X speedup in circuit execution, and an order of magnitude improvement in fidelity, compared to the best-known compilers over a representative set of VQE and QAOA benchmarks. Tetris is open-sourced at this \href{https://anonymous.4open.science/r/vqe_project-3E21/README.md}{link}.


\section{Motivation}
\label{sec:motivation}


The flexibility of circuit synthesis when dealing with Pauli-strings offers a significant opportunity for gate cancellation and the reduction of SWAP gates. In this regard, we make three main observations that can help reduce gate count, reduce circuit depth, and enhance fidelity.

\emph{\textbf{Observation 1 -- Overlooked Potential for 2Q Gate Cancellation}}: The potential for 2-qubit gate cancellation is often underestimated in the compilation of VQA kernels, especially in the variational quantum eigensolver (VQE) kernels. Prior work \cite{li+:asplos22, li+:isca21} schedules similar Pauli-strings over time, which guarantees to maximize 1-qubit gate cancellation but not necessarily 2-qubit gate cancellation.

\begin{figure}[htb]
    \centering
    \begin{subfigure}{0.5\linewidth}
        \centering
        \includegraphics[width=\linewidth]{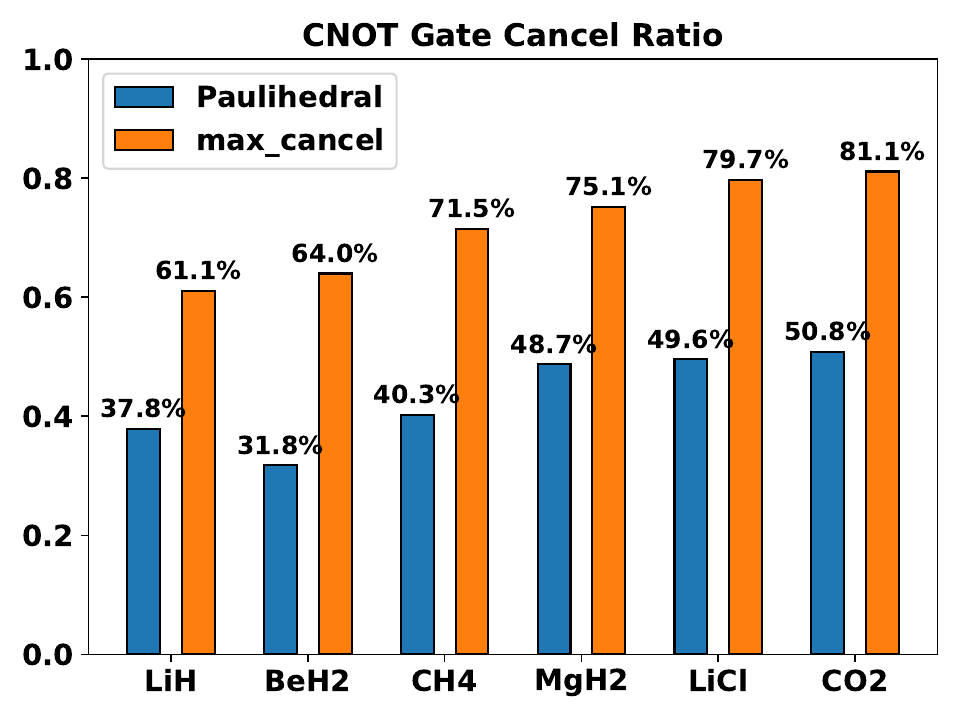}
        \caption{Jordan-Wigner}
        \label{fig:sub1}
    \end{subfigure}
    \hspace{-0.03\linewidth}
    \begin{subfigure}{0.5\linewidth}
        \centering
        \includegraphics[width=\linewidth]{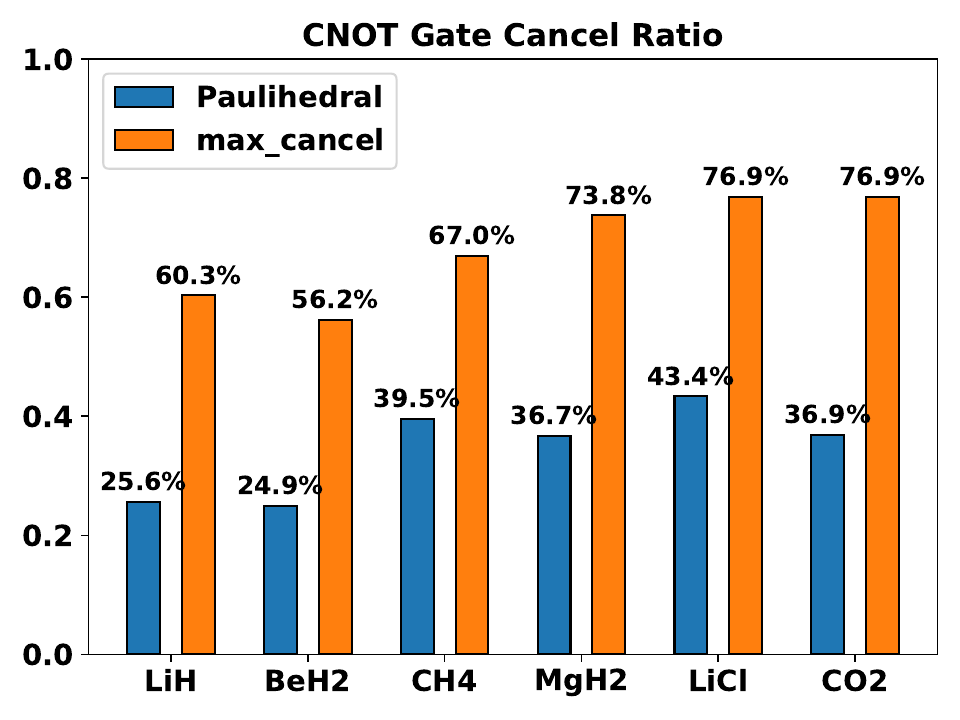}
        \caption{Bravyi-Kitaev}
        \label{fig:sub2}
    \end{subfigure}\vspace{-0.05in}
    \caption{The CNOT gate cancellation opportunities in VQE applications for real molecules.  ``Paulihedral" is by Li \etal \cite{li+:asplos22}. ``max\_cancel" corresponds to the maximum number of CNOT that can be canceled \eddy{with the Pauli-string grouping implied by Jordan-Wigner or Bravyi-Kitaev encoding. We did not apply other circuit optimizations like those in BQSKit \cite{bqskit}.} }
    \label{fig:cancelrealnumbers}
\end{figure}
\vspace{-0.05in}

We compare the maximal percentage of 2-qubit gates that can be canceled and the actual percentage of 2-qubit gates canceled by the state-of-the-art VQE compiler Paulihedral \cite{li+:asplos22} in Fig. \ref{fig:cancelrealnumbers}. It shows the results for six real molecules with 12 to 30 qubits and 640 to 20K Pauli-strings, using the Jordan-Wigner and Bravyi-Kitaev encoding method for the UCCSD ansatz, one of the most studied chemistry-inspired ansatz. How the maximal 2-qubit gate cancellation is estimated will be discussed in Observation 2. It can be seen that the maximum 2-qubit gate cancellation ratio is high across all molecules, from 61.17\% to 81.1\% for the Jordan-Wigner encoder and 60.3\% to 76.9\% for the Bravyi-Kitaev encoder. In comparison, Paulihedral cancels at most 50.8\% for the Jordan-Wigner encoder and 43.4\% for the Bravyi-Kitaev encoder.

\begin{figure}[htb]
    \centering
\includegraphics[width=0.45\textwidth]{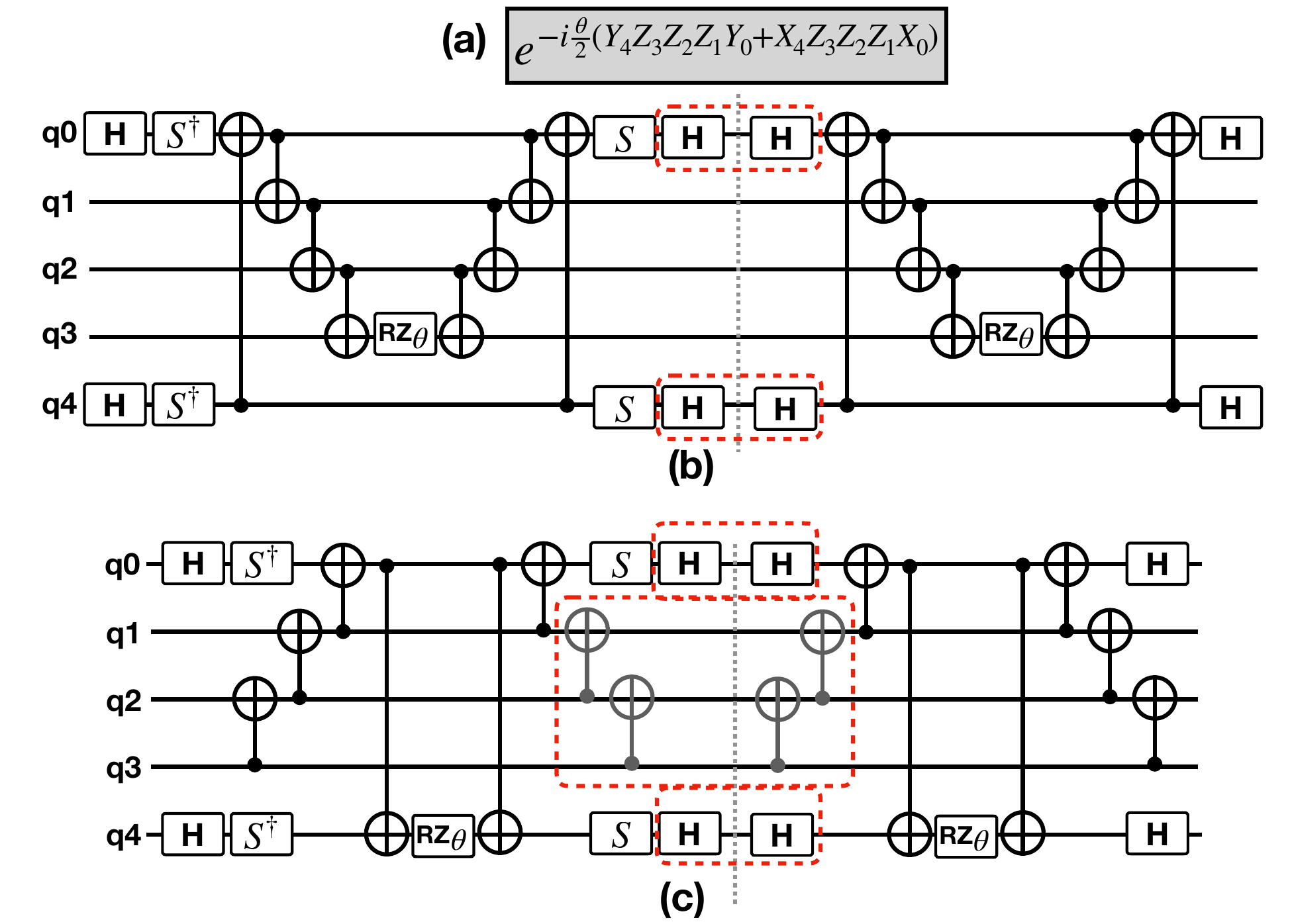}\vspace{-0.05in}
    \caption{Two different circuits lead to different gate-canceling results. (a) Matrix exponential; (b) Only cancels 1Q gates; and (c) Cancels both 1Q and 2Q gates.  }
    \label{fig:cancel_against_coupling}
\end{figure}

\emph{\textbf{Observation 2 -- Gate Cancellation Due to Similarity in Pauli-strings}}: 
Opportunities for gate cancellation exist primarily due to the substantial similarity among Pauli-strings. We show an example in {Fig. \ref{fig:cancel_against_coupling}}. The matrix exponential is $e^{-i\frac{\theta}{2}( Y_0Z_1Z_2Z_3Y_4 + X_0Z_1Z_2Z_3X_4)}$.

For this matrix exponential, two sub-circuits need to be synthesized, corresponding to the two Pauli-strings $Y_0Z_1Z_2Z_3Y_4$ and  $X_0Z_1Z_2Z_3X_4$. Fig. \ref{fig:cancel_against_coupling} {(b)} and (c) show two different ways to synthesize these two sub-circuits, which lead to varying numbers of gates being canceled. Here, two back-to-back Hadamard gates cancel each other, and two back-to-back CNOT gates cancel as well. We can see that Fig. \ref{fig:cancel_against_coupling} {(b)} and (c) cancel the same number of 1-qubit gates. But Fig. \ref{fig:cancel_against_coupling} (c) eliminates 4 CNOT gates, and Fig. \ref{fig:cancel_against_coupling} (b) eliminates none. Everything else is the same (including the initial qubit mapping) for these two circuits, except that they use two different CNOT trees. 

\label{sec:2qubitgateerror}
As a 2-qubit gate has an order of magnitude higher error and execution time than a 1-qubit gate\fixit{\cite{tannu+:asplos19}\cite{IBMQiskit}}, reducing both CNOT and 1-qubit gates is preferable to only 1-qubit gates. 

\begin{figure}[htb]
    \centering
\includegraphics[width=0.35\textwidth]{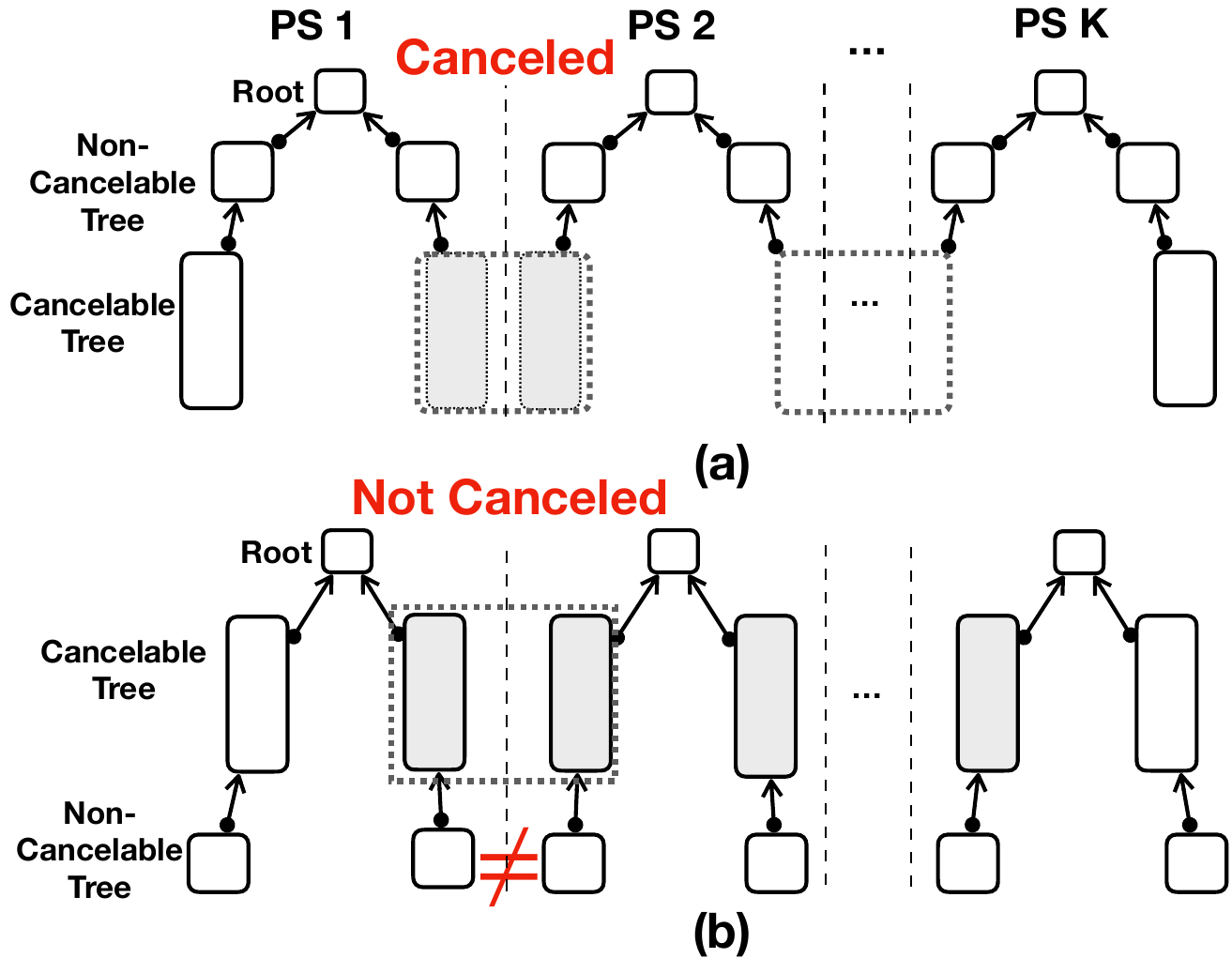}\vspace{-0.05in}
    \caption{The impact of qubit placement in the synthesized tree. (a) places qubits with common operators at the leaf section -- resulting in cancelable tree branches, while (b) does not. This leads to no gate cancellation in (b) despite significant opportunities for doing so. }
    \label{fig:cancel_abstract}
\end{figure}

 In Fig. \ref{fig:cancel_abstract}, if there are cancellable gates corresponding to common Pauli-operators, and if we place the involved qubits at the leaf section of the tree, it allows for the cancellation of all cancellable gates. The leaf section of the tree is the part of tree on the bottom as shown in the cancelable tree in Fig. \ref{fig:cancel_abstract} (a) and the non-cancelable tree in Fig. \ref{fig:cancel_abstract} (b).
 
 In contrast, if the cancellable gates are not placed at the leaf tree section, the presence of non-cancellable gates at the leaf section of the tree can impede this cancellation process, as illustrated in Fig. \ref{fig:cancel_abstract} (b). No matter how many cancellation opportunities exist, no gate can be canceled.  This explains the result in Fig. \ref{fig:cancel_against_coupling} (b) and (c). For Fig. \ref{fig:cancel_against_coupling} (c), the CNOT gates between q1, q2, and q3 are placed in the leaf section of the overall tree; hence, they can be canceled. But in Fig. \ref{fig:cancel_against_coupling} (b), they are placed close to the tree's top section, preventing cancellation.

 We obtain the maximum-cancellation numbers in Fig. \ref{fig:cancelrealnumbers} by placing the subset of qubits that share a maximum number of non-identity operators in the leaf section of the tree, and hence the corresponding logical circuit has maximum CNOT cancellation results.

\emph{\textbf{Observation 3 - Root Cause for Pauli-String Similarity}}:
There is a high similarity between Pauli-strings in current computational chemistry -- the VQE applications. With the popular UCCSD \cite{peruzzo+:nature14} ansatz circuit and the commonly used encoder, Jordan-Wigner \cite{jordan+:zf28}, or Bravyi-Kitaev \cite{bravyi+:ap02}, a Pauli-string is padded with Pauli-Z or Pauli-X operators to enforce Fermionic sign prescription in addition to the four pivotal fermionic-to-spin operators. For instance, for an 18-qubit (orbital) CH4 molecule, the number of padded Z operators can be as much as 14. Hence, two consecutive Pauli-strings show a high similarity. This leads to abundant 1-qubit gate cancellation opportunities, already explored by prior work, and 2-qubit gate cancellation opportunities, \textbf{only if the circuit is synthesized properly.}

\vspace{-6pt}
\fixit{
\section{Comparison to Prior Art}
 \label{subsec:differencewithPH}
 \vspace{-2pt}
 Paulihedral \cite{li+:asplos22} exploits 1-qubit gate cancellation and places emphasis on SWAP reduction. Paulihedral finds the maximum connected component (CC) for already mapped qubits on the hardware and then grows the tree from the CC. This method results in a SWAP-centric circuit synthesis. }

\fixit{The difference is that our work focuses on both 2-qubit gate canceling and SWAP reduction, while Paulihedral only focuses on SWAP reduction. In terms of tree construction, our work specifies which qubits must be in the root-tree-section (the part has a lower depth), and which qubits must be in the leaf-tree-section (the part that has a larger depth than the root-tree section). However, Paulihedral allows any qubit to be the root or leaf, as long as maximum SWAP reduction is achieved, and hence might miss 2-qubit gate canceling opportunities. }

\begin{figure*}[htb]
    \centering
\includegraphics[width=1\textwidth]{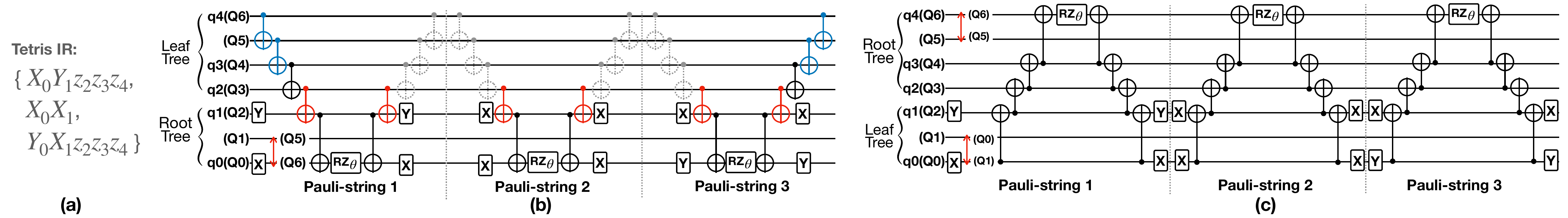}
    \caption{Tetris-IR and the generated circuit. (a) Tetris-IR, where we omit the weights and common rotation angles. (b) Setting q0 and q1 as the root tree.  (c) Setting q2, q3, and q4 as the root tree. (b) has more cancellations than (c). Assuming the underlying hardware is line connected from Q0 to Q6. Initial mapping is indicated, i.e., q4(Q6) means q4 mapped to Q6.}\vspace{-0.1in}
    \label{fig:new_pauli_IR}
\end{figure*}

\section{Our Proposed Solution and Design Tradeoffs}

We propose the \textit{Tetris} compiler and the \textit{Tetris}-IR.  The Tetris compiler has the capability of exploiting 2-qubit gate cancellation while simultaneously minimizing the cost of circuit compilation. The Tetris-IR is a refined intermediate representation that symbolizes a group of Pauli strings sharing common Pauli operators, which encodes the 2-qubit gate canceling potential.

\subsection{Distinction Between the Root Tree and Leaf Tree Qubit Set}

Tetris makes the distinction between two types of qubits: the \emph{root-tree-qubit-set} and the \emph{leaf-tree-qubit-set}.

Before describing the two sets, we define the concept of Tetris blocks. Prior work \cite{li+:asplos22, li+:isca21} uses the abstraction of Pauli-strings to coordinate the optimizations at different technology stacks. An IR of Pauli-strings is broken down into a list of Pauli-string blocks. Each block consists of Pauli-strings that share a factor of rotation angle. The fact that Pauli-strings are broken down into blocks is relevant to how the ansatz is constructed \cite{tilly+:physicsreports22}, but irrelevant to this paper. What is relevant to this paper is that Pauli-strings within a block have relatively higher similarity, and Pauli-strings across different blocks may still have high similarity but are lower than the ones within a block. We define a Tetris block as a block from the ansatz construction process. A Tetris block could also consist of multiple blocks from the ansatz construction process. Our methodology will apply, too.


The \textbf{leaf tree qubit set} is the maximum qubit set over which the corresponding Pauli-operators are the same for all strings in a Tetris block. The \textbf{root tree qubit set} comprises the rest of the qubits which correspond to non-identity operators.

The leaf tree(s) must point to the root tree. Since the leaf-tree qubit set contains the qubits that share Pauli operators, if the leaf trees have the same structure across multiple Pauli-strings, all 2-qubit gates between qubits in the leaf-tree qubit set can be canceled, except the two leaf trees that are at the first Pauli-string and the last Pauli-string, as shown in Fig. \ref{fig:cancel_abstract}. 

Tetris allows flexibilities in constructing the root tree and the leaf tree(s), but Tetris enforces that the root-tree-qubit-set must be used to build the root tree only, and the leaf-tree-qubit-set must be used to build the leaf tree(s) only.

An example of a block of three Pauli-strings is shown in Fig. \ref{fig:new_pauli_IR} {(a)}, where qubits q2, q3, and q4 have three Z operators in common; hence, they belong to the leaf-tree-qubit-set. The root-tree-qubit-set comprises qubits q0 and q1 in Fig. \ref{fig:new_pauli_IR} (b). 

 As opposed to Tetris, 
 a possible synthesized circuit is shown in Fig.\ref{fig:new_pauli_IR} {(c)}, where the non-cancellable gates in the lowest level of the tree prevent the cancellation opportunities, and hence has 8 more CNOT gates than the circuit in Fig.  \ref{fig:new_pauli_IR} {(b)}.

Note that the CNOT(q4, q3) in Fig. \ref{fig:new_pauli_IR} (b) is replaced by two CNOT gates using the bridging approach, shown in blue. We will discuss the details of fast bridging in Section \ref{sec:fastbridge}.

\subsection{Tetris IR and Tetris Tuning Spectrum}

We introduce Tetris IR, which allows flexibility of circuit synthesis and, in the meantime, exposes 2-qubit-gate-cancellation opportunities.

\subsubsection{Tetris IR} Tetris IR is a refined IR of Pauli-strings. It is designed for the compiler to capture the information of the root-tree-qubit-set and the leaf-tree-qubit set, allowing the compiler to explore the tradeoff between 2-qubit-gate-cancellation and hardware-aware SWAP reduction.

\begin{figure}[htb]
    \centering
    \includegraphics[width=0.4\textwidth]{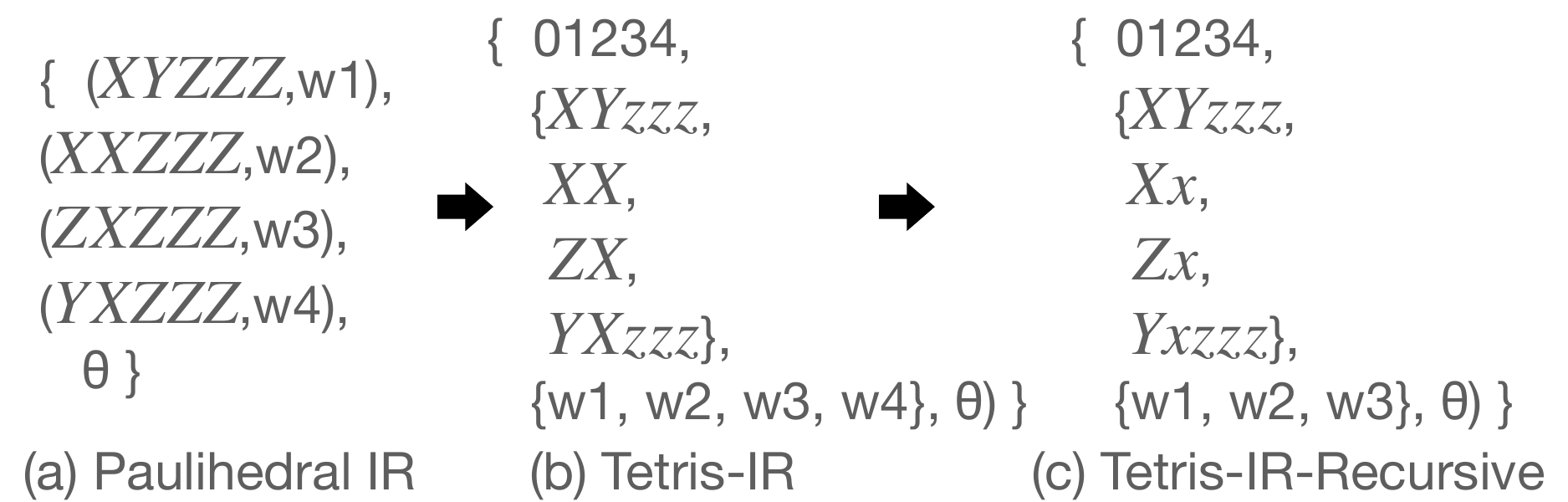}
    \caption{Tetris-IR}
    \label{fig:tetrisIR}
\end{figure}

 A Tetris IR is also a list of Pauli-string blocks, similar to that in prior work \cite{li+:asplos22, li+:isca21}. In the original Pauli-string IR, it also needs to indicate the weight and the common factor of rotation angles or time steps. An example of a Tetris block is shown in Fig. \ref{fig:tetrisIR} (b). In Tetris IR, each block starts with an integer sequence annotating the order of the qubits after regrouping into the root-tree-qubit-set and leaf-tree-qubit-set. Then, each string starts with a non-common section of the Pauli-string. Only the first and last string has the common section of the Pauli-strings. We also make the common section's Pauli operators lower case in Fig. \ref{fig:tetrisIR} (b); since only the peripheral section of the tree for the common Pauli operators will be kept, the middle leaf sections for these Pauli operators can be canceled out. To show how it is different from prior work, we also show the Paulihedral IR representation in Fig. \ref{fig:tetrisIR} (a). 

 The weights $w1$ to $w3$ and the angle $\theta$ do not impact the circuit structure. We still keep them in the IR but do not discuss them because they are not relevant to this paper.

There is also a version of IR called Tetris-IR-recursive shown in Fig. \ref{fig:tetrisIR} (c). It recursively deletes common sub-strings and makes the corresponding peripheral tree operators small-case. For instance, the last two Pauli-strings also have a cancellation opportunity for the second qubit with the Pauli-X operator. They correspond to the cancellable X and CNOT gates in Fig. \ref{fig:new_pauli_IR} (b) and (c). In our paper, we will exploit Tetris-IR, not the recursive one. But we leave it as our future work.

\subsubsection{Tetris Tuning Spectrum}
\label{sec:tetrisSpectrum}

Different circuits can be synthesized from a Tetris-IR. Specifically, the compiler can adaptively tune the number of 2-qubit gates that are eliminated.

One extreme end of the Tetris tuning spectrum is to prioritize 2-qubit gate cancellation. That is, to cancel all 2-qubit gates between the qubits with common operators. 
In this case, it is necessary to construct a single leaf tree connecting all qubits assigned to the leaf-tree-qubit-set. 

\begin{figure}[htb]
    \centering
\includegraphics[width=0.42\textwidth]{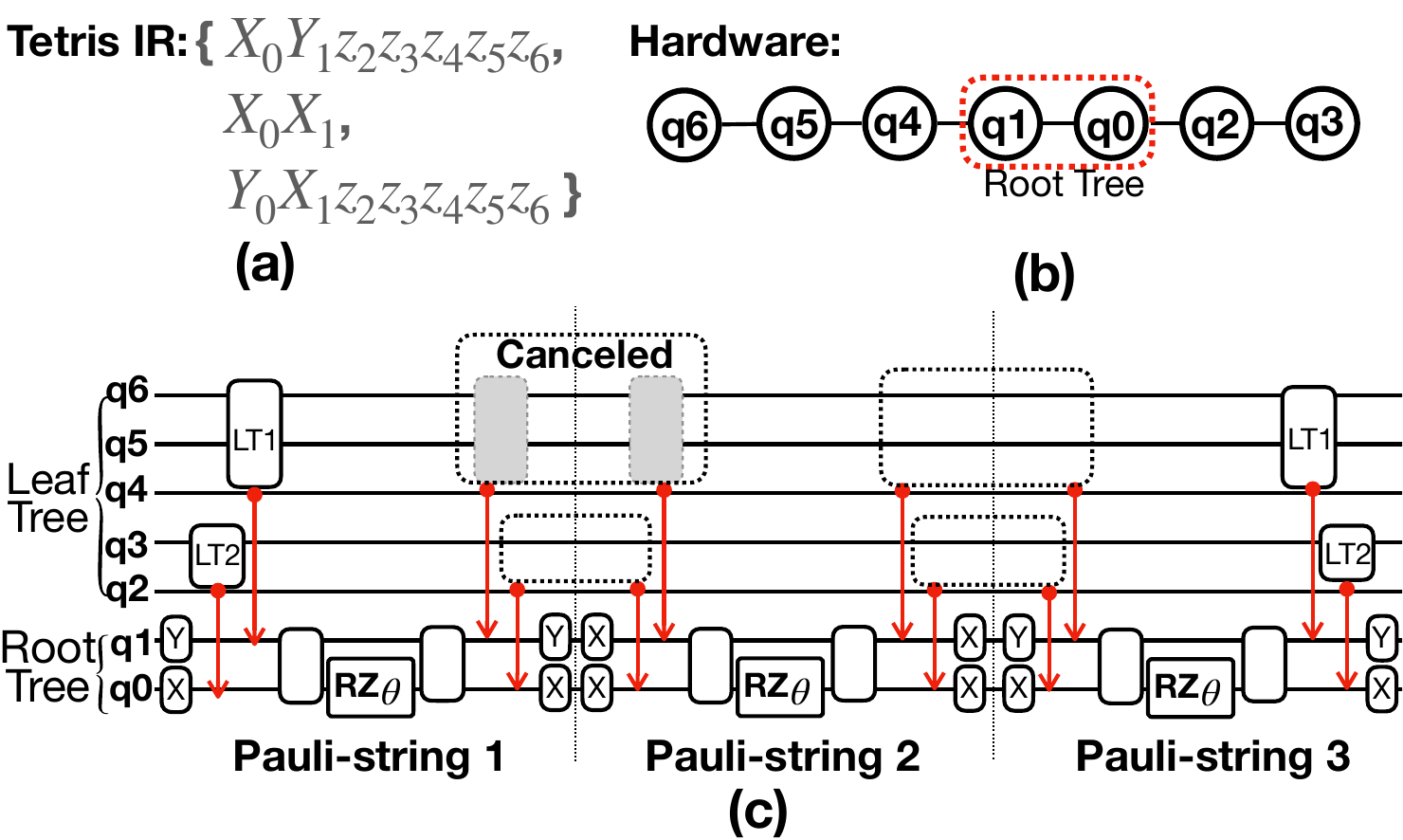}\vspace{-0.05in}
    \caption{Tetris adaptively tuning 2-qubit gate cancellation. There are two leaf trees here. The leaf tree \{q4, q5, q6\} pointing to q1 and the leaf tree \{q2, q3\} pointing to q0. The gates in each leaf tree are canceled and yet there is no SWAP needed.}
    \label{fig:tunableTetris}
\end{figure}

However, maximum canceling might incur unnecessary SWAP insertion overhead. As the example shows in Fig. \ref{fig:tunableTetris}, the root-tree-qubit-set \{q0, q1\} is already mapped to the middle in the hardware coupling graph in  Fig. \ref{fig:tunableTetris} (c) before the Pauli-string is converted to a sub-circuit. To construct a single tree that connects all qubits assigned to the leaf tree set \{q2, q3, q4, q5, and q6\}, at least 4 SWAP gates are needed (i.e., moving q2 towards q4, and q3 towards q4). The single leaf tree here will have 16 CNOT gates canceled from synthesis. 

On the other hand, if we do not require the maximum number of 2-qubit gates canceled, we can allow more than one leaf tree. If we allow the qubits in the leaf-tree-qubit-set to form multiple leaf trees connected to the root tree, as shown in Fig.  \ref{fig:tunableTetris} (c), we still have 12 CNOT gates canceled without inserting extra SWAP gates. Compared with the extreme case of a maximum 2-qubit gate canceling setup, this setup of having 2 leaf trees for each Pauli-string, has 4 fewer CNOT gates (12 versus 16) canceled but also saves 4 SWAP gates (which is 12 CNOT gates). Hence, the Tetris-IR is flexible in that it only specifies the root tree qubit set and the leaf tree qubit set, respectively. It does not specify how many leaf trees are allowed or how the root/tree trees should be constructed.

Tetris compiler adaptively tunes the level of 2-qubit gate cancelation and SWAP insertion, as discussed in Section \ref{sec:tech}. 


\subsection{SWAP v.s. Fast Bridging} 
\label{sec:fastbridge} 
\begin{figure}[htb]
    \centering
    \includegraphics[width = 0.45\textwidth]{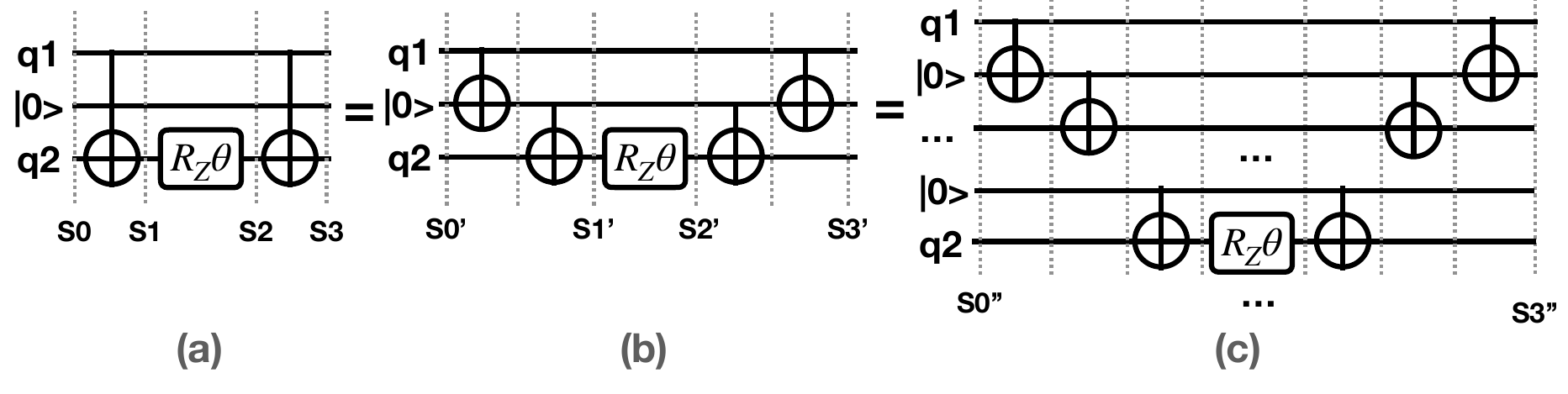}\vspace{-0.05in}
    \caption{The fast CNOT bridge using ancilla qubit(s) in state $|0\rangle$. (a) Original circuit, (b) using one qubit as the bridging qubit, and (c) using multiple qubits as the bridging qubits.}
    \label{fig:fast_bridge}
\end{figure}

In certain cases, there may not be enough similarity among Pauli-strings. For instance, in QAOA, for each Pauli-string, there are at most 2 qubits that have a non-identity operator. Hence, there is not much 2-qubit gate cancellation opportunity. For this type of VQA application, we propose the \emph{fast bridging} approach to help the routing. 
\label{sec:qubitreuse}
The fast bridging method uses extra qubits that are initialized to $|0\rangle$ as bridges to avoid SWAP gates\fixit{\cite{hua+:asplos2023}}. We show examples in Fig. \ref{fig:fast_bridge}. In this example, assuming the underlying hardware coupling is linear, to enable CNOT gates between q1 and q2 in Fig. \ref{fig:fast_bridge} (a), one needs a SWAP gate to move q1 and q2 close to each other. Here, if the ancilla qubit is $|0\rangle$, it can be proved that using the ancilla qubit as a bridge point with 2 CNOT gates shown in Fig. \ref{fig:fast_bridge} (b), the CNOT between q1 and q2 can be achieved.

 Note that, in order to revert the ancilla qubit back to $|0\rangle$, another CNOT between q0 and $|0\rangle$ needs to be inserted. Since we typically have mirrored CNOT gates surrounding the single-qubit gate in the root qubit, we do not have to immediately revert the ancilla qubit back to $|0\rangle$. Only after the second CNOT gate is done we can revert it back to $|0\rangle$, as shown in Fig. \ref{fig:fast_bridge} (b). The fast bridging idea can be extended to multiple ancilla qubits, as shown in Fig. \ref{fig:fast_bridge} (c).  

\begin{figure}
    \centering
\includegraphics[width=0.45\textwidth]{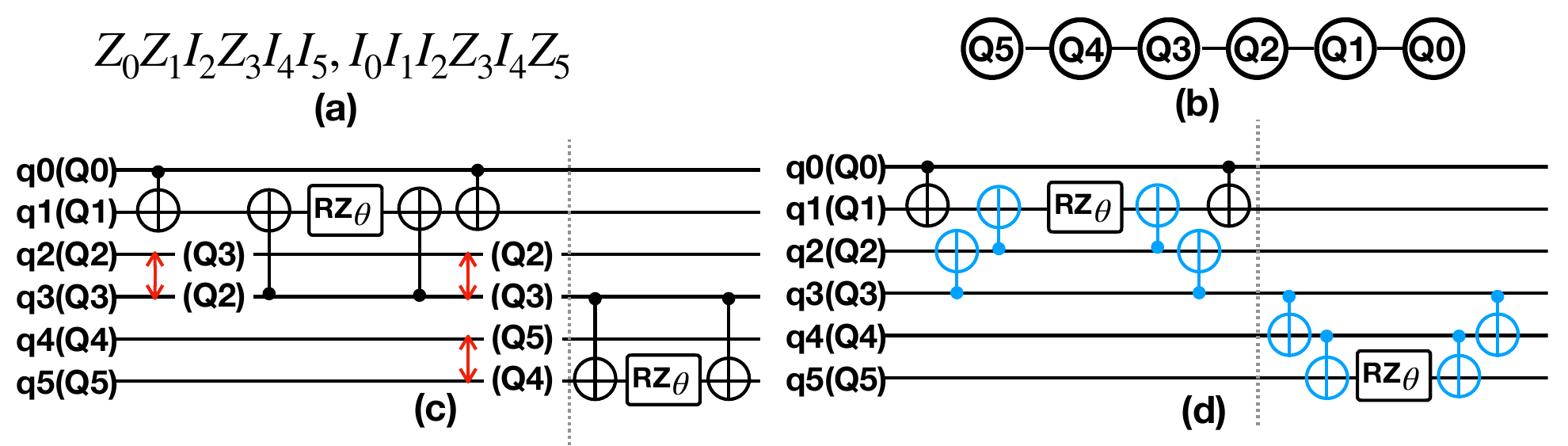}\vspace{-0.05in}
    \caption{Fast Bridging Method is better than SWAP Insertion Method. (a) Two Pauli-strings. (b) The underlying hardware. (c), (d) The compiled circuit of the input Pauli-strings using the SWAP insertion method and CNOT bridge method. }
    \label{fig:bridgeBetter}
\end{figure}

\begin{figure}
    \centering
\includegraphics[width=0.3\textwidth]{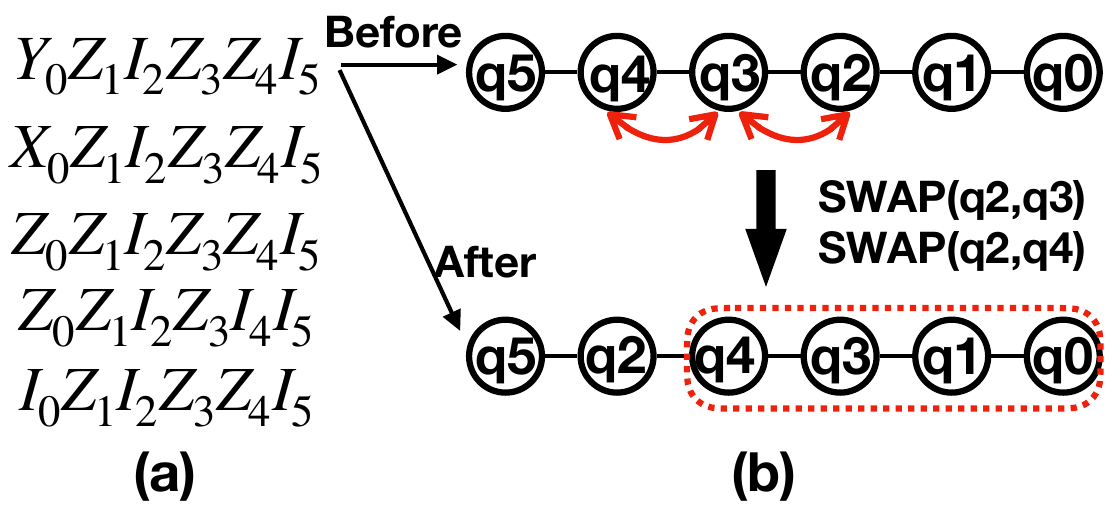}\vspace{-0.05in}
    \caption{Fast Bridging Method worse than SWAP Insertion Method. (a) Five Pauli-strings. (b) The underlying hardware with the qubit mapping before and after implementing the first Pauli-string. All data qubits are connected now.}
    \label{fig:swapBetter}
\end{figure}

We can adaptively choose between fast bridging and SWAP insertion. If a qubit mapping is reused multiple times, we want to prioritize SWAP insertion over bridging.  As shown in Fig. \ref{fig:swapBetter}, two SWAP gates required to implement the first Pauli-string resolve the hardware constraints for all subsequent Pauli-strings. If a qubit mapping does not help further gates by some lookahead, we may prioritize bridging over SWAP insertion. An example is shown in Fig. \ref{fig:bridgeBetter}. 

 The fast bridging method needs to have available ancilla qubits.  The idle qubits utilized for the bridge must be in state $|0\rangle$.  We use mid-circuit measurement to reclaim qubits. Note that it was discovered by Hua \etal \cite{hua+:asplos2023}, that there are substantial mid-circuit measurement opportunities for QAOA circuits.  For fast bridging, after mid-measured qubits are reset to $|0\rangle$, they can be reused in the future. 

While the fast bridging method is designed for the case when there is not much similarity between Pauli-strings, it is also useful for the case where we have a lot of similarities. We use the fast bridging method for the leaf-tree-qubit-set qubits, and SWAPs for the root-tree-qubit-set qubits since the leaf-tree gates run less frequently than the root-tree gates. See our prior example in Fig. \ref{fig:new_pauli_IR} (b).

\section{Technical Implementation}
\label{sec:tech}

\begin{figure}[htb]
    \centering
    \includegraphics[width=0.3\textwidth]{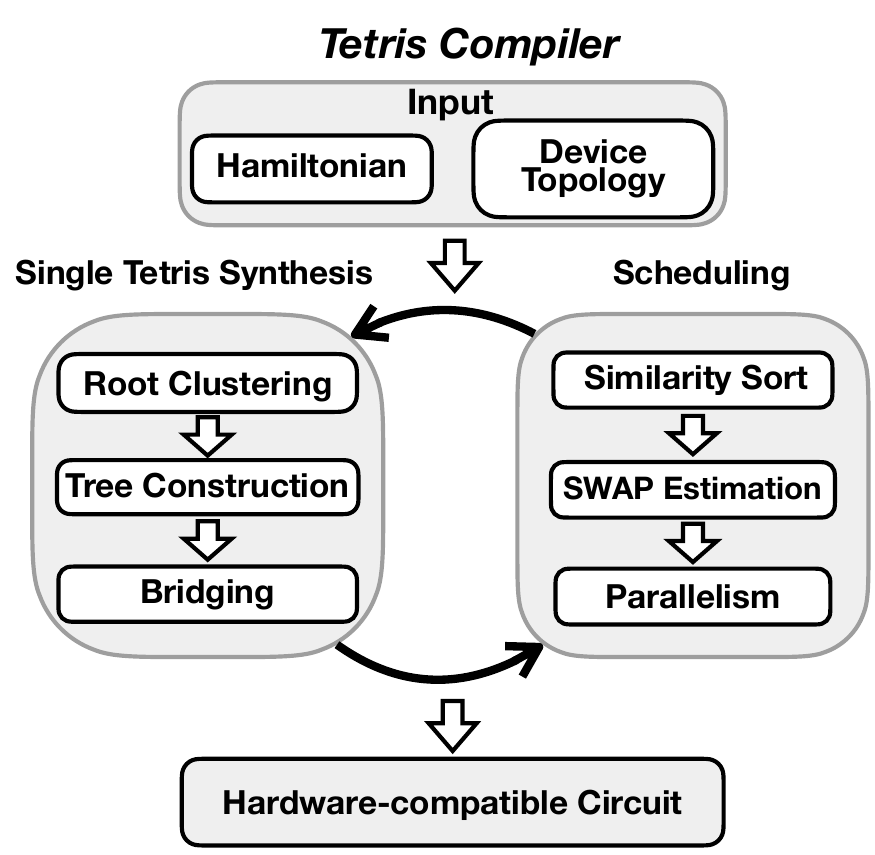}
    \caption{Overview of the Tetris compiler.}
    \label{fig:frame}
\end{figure}

To strike a balance between gate cancellation and the compilation cost for hardware-compliant gates, we propose a compilation framework called ``Tetris," as depicted in Fig. \ref{fig:frame}. In this section, we discuss the key components of Tetris, specifically how to insert SWAP gates to move qubits in a clustered manner. We also present a scheduling strategy to mitigate the cost between two Tetris blocks.

\subsection{Circuit Synthesis with Respect to Hardware}
\label{sec:syntech}

\begin{algorithm}
\begin{algorithmic}[1]
\caption{Circuit Synthesis}
\label{alg:Synthesis}
\Procedure{$Circuit\_Synthesis$}{$G_{coupling}$, $\pi$, $q^{r}$,$q^{l}$}
\State $circuit$ = \{\}
\State $tree$ = \{\}

\State $center$, $paths$ = findCenter($G_{coupling}$, $\pi$, $q^{r}$)
\For {$q^{r}_{i}$ $\in$ $q^{r}$}
    \State circuit.insert(SWAP($q^{r}_{i}$, paths.parent($q^{r}_{i}$)))
    \State tree.insert($q^{r}_{i}$)
\EndFor
\While {$q^{l}\neq \emptyset$}

\State $q^l_j,q_{m}$ = $\underset{q^{l}_{j} \in q^{l},q_m \in tree}{\text{argmin}}$(getScore($q^{l}_{j}$, $q_m$, $w$))
    \State circuit.insert(SWAP($q^{l}_{j}$, paths.parent($q_{m}$)))
    \State tree.insert($q^{l}_{j}$)
    \State $q^l$.remove($q^{l}_{j}$)
\EndWhile
\For{leaf in tree}
    \State $circuit$.insert(CNOT(leaf, tree(leaf).getparent))
\EndFor
\State \textbf{return} $circuit$
\EndProcedure
\end{algorithmic}
\end{algorithm}

\eddy{The circuit is synthesized on the fly. The structure of the circuit synthesis tree is not fixed at the beginning. We will first map the root-tree group's qubits such that they are connected. Then, we perform mapping for each qubit in the leaf-tree group. We process one such qubit at one time, by how close they are to existing mapped qubits. While the qubits are being mapped, the synthesized tree is gradually generated. }


\fixit{For the efficient mapping of all root-tree qubits, we first identify a central point among the mapped qubits $q^{r}_{i}$ (in the root-tree-qubit-set) in the hardware coupling graph, $G_{coupling}$. We let $q^{r}$ denote the set of qubits in the root tree. }

\begin{figure}[htb]
    \centering
\includegraphics[width=0.4\textwidth]{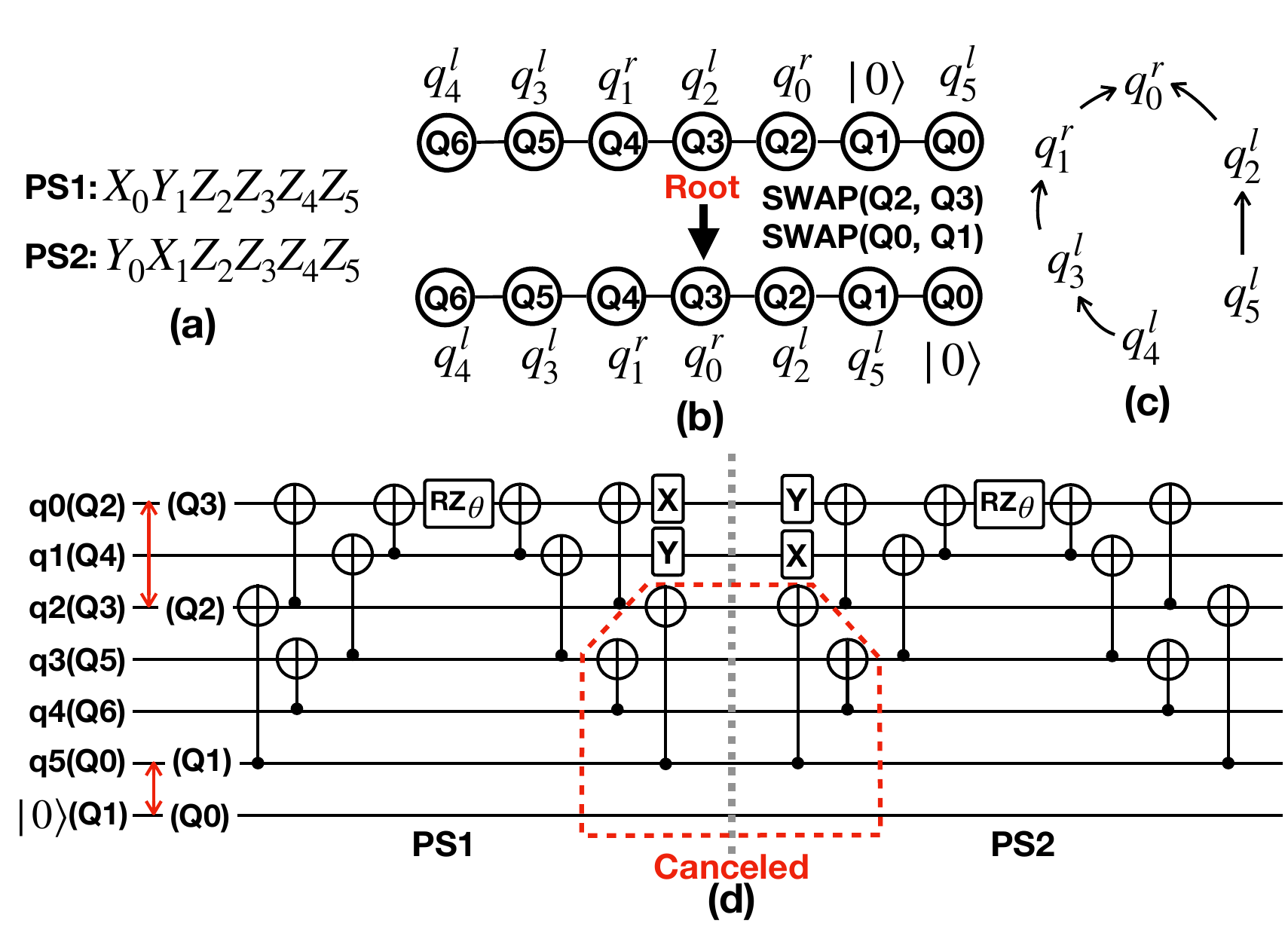}
    \caption{\fixit{ (a) Two Pauli-strings. (b) The coupling graph and qubit mapping. Physical qubit Q3 is selected as the center for the root tree. (c) One of the logical circuit synthesis trees with q0 as the root qubit for both Pauli-strings. (d) The detail of the whole compiled circuit. Four CNOT gates are canceled.  }}
    \label{fig:method_example}
\end{figure}

\fixit{Subsequently, we introduce SWAP gates to cluster all qubits $q^{r}_{i}$ around this center. It's worth noting that the central point doesn't necessarily map to some root-tree qubit $q^{r}_{i}$ at the beginning. A root-tree qubit will eventually be swapped into this position. For instance, in Fig. \ref{fig:method_example}(b), we can employ a SWAP(Q2, Q3) gate to ensure  $q_0^{r}$ is mapped to the central node. We process root-tree qubits one at a time based on their proximity to the already mapped root-tree qubits, until all root-tree qubits are connected in the hardware coupling graph. }

\fixit{Next, we move the qubits $q^{l}_{j}$ in the leaf-tree-qubit-set based on their proximity to already mapped nodes. We also determine the parent node of each leaf-tree-qubit in this process. The details are described in the next section. }

\fixit{ We show a concrete example in Fig. \ref{fig:method_example}.  The corresponding pseudocode for the whole process is detailed in Algorithm \ref{alg:Synthesis}.}


\paragraph{Circuit Synthesis and SWAP Cost Control} As discussed in Section \ref{sec:tetrisSpectrum}, gathering all leaf tree qubits into a single leaf tree may cause a lot of SWAP gates overhead and diminish the benefit from gate cancelation. We designed a cost function to achieve a good trade-off between gate cancellation and SWAP cost. The cost function for moving a leaf-tree qubit $q_n$ close to another qubit $q_m$, and marking this qubit $q_m$ as its parent is below:

\[
    score(q_n, q_m,w) = (d-1)*w +
\begin{cases}
    \#ps*2 ,& \text{if } \fixit{q_{m}} \in q^{r}
    \\
    2,              & \text{otherwise}
\end{cases}
\] 

$q_{n}$ $\in$ $q^{l}$ and $q_{n}$ is not mapped yet. $q_{m}$ is a root tree qubit or a mapped leaf tree qubit. $\#ps$ is the number of Pauli-strings in a Tetris block. $d$ is the shortest distance between $q_{n}$ and $q_{m}$ \eddy{(When we search for the shortest path, we remove the already mapped nodes except $q_m$ itself, such that already mapped qubits are not affected.)}. $w$ is the weight for SWAP cost. \eddy{We can tune the weight $w$. If the weight $w$ is set to high, the compiler will try to move the unmapped qubit to the nearest mapped qubit to reduce the SWAP gate count but potentially miss a gate cancelation opportunity. If the weight is low, the compiler will favor connecting the unmapped qubit to a leaf tree qubit that maximizes 2-qubit gate cancelation.} \eddy{In our experiment, we tested different $w$ values ({detailed results are added in the evaluation}) and found that $w$ = 3 happens to be a good choice among other good choices. At the same time, 3 corresponds to the fact that one SWAP consists of three CNOT gates. }

 If $q_{m}$ is a root tree qubit, then in addition to the SWAP cost, the score needs to include all (not canceled logical) CNOT gates between $q_{n}$ and $q_{m}$. Otherwise, the CNOT gates can be canceled except at the Tetris unit's beginning and end.  
We calculate the score of connecting $q_{n}$ with each root tree qubit or mapped leaf tree qubit and find the one with the minimum cost. Then, we insert SWAP gates based on the shortest distance path found. This process continues until all $q^{l}$ qubits are mapped.  

\begin{figure}[htb]
    \centering
\includegraphics[width=0.4\textwidth]{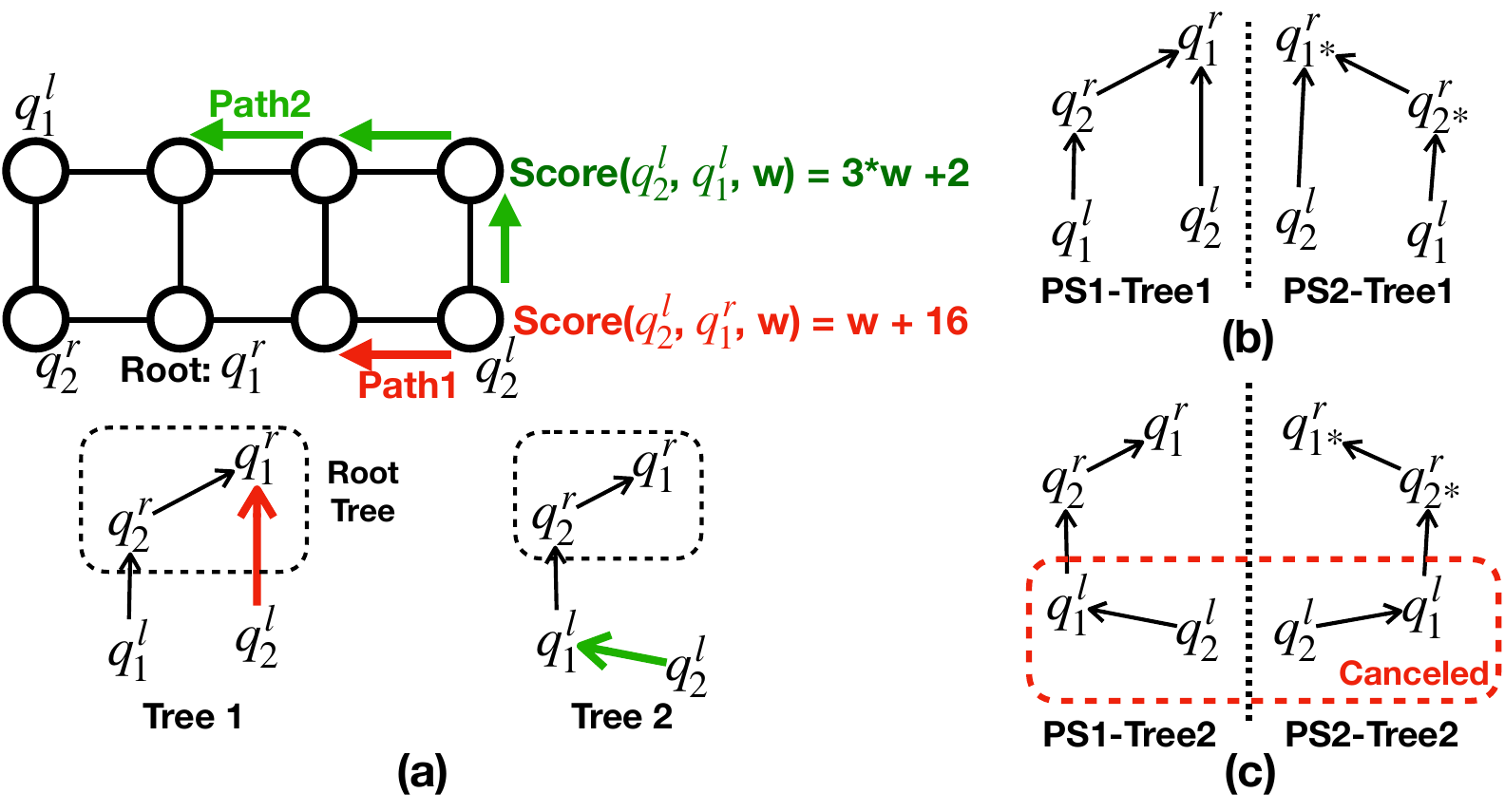}
    \caption{ \eddy{SWAP Weight $w$ Analysis. (a) Two circuit synthesis trees. Tree 1 has two leaf trees and a lower SWAP cost $w$, and Tree 2 has one leaf tree and a higher SWAP cost $3w$. (b) There is no gate cancelation for Tree 1. (c) Two CNOT gates are canceled for Tree 2. Note that PS1 and PS2 have the same root tree qubit set but different operators on the root tree qubit set by our definition of root-tree-qubit-set.}}
    \label{fig:costFun_example}
\end{figure}

\fixit{For instance, there are two valid circuit synthesis choices in Fig. \ref{fig:costFun_example}. Circuit synthesis tree 1 tries to reduce the SWAP gate count, and circuit synthesis tree 2 focuses on gate cancelation. The score calculated based on tree 1 is $w + 16$. Because linking $q^{l}_{2}$ to $q^{r}_{1}$ requires one SWAP gate and would not contribute to the gate cancelation (need \#ps*2 = 16 CNOT gates if we let \#ps=8). For connecting $q^{l}_{2}$ and $q^{l}_{1}$, we need at least 3 SWAP gates but achieve a cancelation of 14 CNOT gates and the CNOT cost is $16-14=2$  {(if weight $w$ is set to 3).}
Tetris would favor tree 2 synthesis since the cost of tree 2 is $3*w+2=11$, while that of tree 1 is $w+16=19$.
We could also tune the SWAP weight $w$ to 8, favoring tree 1's synthesis.
}

\paragraph{Bridge insertion} It is considered in the process of the circuit synthesis for the leaf tree. As we discussed in Section \ref{sec:fastbridge}, the fast bridging method has the advantage over the SWAP insertion method only when a SWAP gate does not benefit multiple future CNOT gates. In the process of circuit synthesis for the leaf tree, we already decided on the routing paths that $q^{l}$ will move along with. In these paths, if existing qubits are in state $|0\rangle$, we consider that qubits have a Pauli operator Z on them. Such that we would apply CNOT gates on those ancilla qubits. \fixit{ For example, in Fig. \ref{fig:method_example}(d), instead of inserting SWAP(Q0, Q1) for CNOT(q5, q2), we could bridge q5, q2 by using ancilla qubit Q1 with CNOT(q5, Q1), CNOT(Q1, q2). These two gates between two Pauli-strings would also be canceled and save one CNOT gate compared to the SWAP insertion strategy.}


\subsection{Block Scheduling}
\label{sec:blockscheduling}
Gate cancellation may also happen between two Tetris blocks. In order to further improve the gate cancellation ratio, it is natural to schedule two similar Tetris blocks adjacent to each other.  For example, the Tetris block in the following has qubits \{1, 6\} in the root tree and qubits \{2-5\} in the leaf tree. It is favored to be scheduled after the Tetris block mentioned in Fig. \ref{fig:tunableTetris}. Because they have four Pauli operators in common in the same sub-set of qubits \{2-5\} from both leaf trees. The qubit q0 with the Pauli-I (identity) operator is ignored. 

\vspace{-10pt}
\begin{align*}
\text{PS1:} &\quad I_0Y_1Z_2Z_3Z_4Z_5X_6, \\
\text{PS2:} &\quad I_0X_1Z_2Z_3Z_4Z_5Y_6, \\
\text{PS3:} &\quad I_0Y_1Z_2Z_3Z_4Z_5X_6.
\end{align*}

\paragraph{Tetris similarity}
We define the similarity between two Tetris blocks by counting the number of common Pauli operators between the leaf trees of two Tetris blocks.
Let $T_1$ and $T_2$ be two Tetris blocks with their respective leaf trees denoted as $LT_1$ and $LT_2$. We define the similarity ($S$) between $T_1$ and $T_2$ based on the size of the common part of their leaf trees as follows:

\begin{equation}
\label{eq:similarity}
S(T_1, T_2) = \frac{|C(LT_1, LT_2)|}{|LT_1| + |LT_2| - |C(LT_1, LT_2)|}
\end{equation}

Where: $|C(LT_1, LT_2)|$  is the size of the common part of  $LT_1$ and  $LT_2$.

This formula quantifies the similarity between two Tetris blocks based on the size of their shared leaf tree components.

\paragraph{Trade-off}As introduced in Sec. \ref{sec:syntech}, it is necessary to insert SWAP gates to cluster root tree qubits. If we only schedule Tetris blocks with the consideration of similarity, we may suffer from the SWAP cost between two Tetris. For the Tetris block mentioned above, if we schedule it immediately after the Tetris block in Fig. \ref{fig:tunableTetris}, then at least two SWAP gates are inserted for the root tree of the second Tetris block.

As we can see here, the intra-block optimizations might hurt inter-block optimizations if scheduling is not done properly. So, a smart scheduler is desired. It has to consider the SWAP cost between two Tetris blocks since the root tree qubits vary during the whole VQE procedure and incur a tremendous SWAP cost in gathering the root tree qubits. The scheduler is also expected to catch the gate cancellation opportunities between two consecutive Tetris blocks. 
\paragraph{Scheduling}The major steps of scheduling are below:

\begin{enumerate}
    \item Sort the blocks by active length in descending order and implement the one with the largest active length. 
    
    \item Sort all the remaining Tetris blocks by the similarity as defined in Eq. \ref{eq:similarity}.
    \item Let the set of blocks of top-K similarity be the candidate blocks. Checking the SWAP cost of grouping root trees for the candidate blocks. Then, we schedule the one Tetris block with the minimum SWAP cost. 
    \item Repeat steps 2-3 until all Tetris blocks are scheduled.
\end{enumerate}

We start with the Tetris block with the longest active length because these Tetris blocks have more gate cancelation opportunities and are more important. The active length means the number of non-identity Pauli operators. In steps 2-3, we schedule the Tetris blocks with the consideration of both gate cancelation and  SWAP cost reduction. Meanwhile, sorting Tetris blocks by similarity would also prioritize the Tetris with higher active length.

\subsection{Bridging Optimization for QAOA} We mentioned in Section \ref{sec:syntech} how bridge is inserted for Pauli-strings with high similarity. We also perform bridging optimization for QAOA circuits, with each Pauli-string having only two non-identity operators. When a choice of SWAP versus \emph{fast bridging} has to be made during the hardware mapping process, we use lookahead. If the SWAP insertion can help future CNOT gates, we insert SWAP, otherwise, we use fast bridge. We also schedule the two-qubit gates to improve the mid-circuit measurement opportunities such that fast bridge can be applied using the reset qubits.

\section{Evaluation}


\subsection{Experiment Setup}
\paragraph{Backend} 
The compiler framework introduced in this paper focuses on near-term superconducting backends, specifically targeting IBM's heavy-hex architecture and {Google Sycamore}.   The target IBM quantum hardware is IBM\_ithaca, with a 65-qubit heavy hexagon structured coupling map. The Sycamore hardware coupling graph is set to 64 qubits with 8 qubits in each row.

\paragraph{Metrics} 
\label{subsec:metrics}
We evaluate the effectiveness of our framework using the following metrics: circuit depth, CNOT gate count, circuit duration, gate cancellation ratio (GCR), circuit compilation time, and {the circuit fidelity (measured through IBM random benchmarking {\cite{magesan+:prl11}})}. Those metrics are widely used in various quantum circuit compilers  \cite{zhang+:arxiv2020, li+:asplos22, zhang+:asplos21, tan+:ICCAD20, molavi+:micro22satmap}.
\emph{Circuit depth} refers to the length of the critical path in the compiled circuit, correlating with the overall duration of the circuit. It's preferable to have a smaller circuit depth, as this can help minimize decoherence errors. {When calculating circuit depth, we break down a SWAP gate into three CNOT gates.} \emph{The CNOT gate count} is the total number of CNOT gates in the compiled circuit, encompassing both the original circuit gates and those decomposed from the added SWAP gates. Since two-qubit gate errors are prevalent in accumulated gate errors, fewer gate counts signify fewer accumulated errors. \emph{Circuit duration} represents the time of the circuit in the unit $dt$ (given by the Qiskit pulse simulator). 

We also showcase the ratio of original circuit logical CNOT gates that have been canceled to demonstrate the effectiveness of our optimization passes. The cancellation ratio is calculated by the following: 
\begin{equation}
    ratio = \frac{canceled\_CNOT\_gate\_count}{original\_circuit\_CNOT\_gate\_count} 
    \label{eq:ratioCalculate}
\end{equation}


\paragraph{Baselines}
\label{subsec:baselines}
 {We primarily compare our method with the following frameworks: Paulihedral (PH) \cite{li+:asplos22}, PCOAST\cite{paykin2023pcoast}, and the general compiler T$|$Ket$\rangle$ \cite{tket}. 
 PCOAST is implemented by Intel® Quantum SDK\cite{khalate2022llvmbased} with Intel’s native gate set \{RXY, CZ\}. For fairness reasons, we convert the output circuits of PCOAST to the circuits using the same basis gate set \{U3, CNOT\} as other methods.}
An additional baseline is the $max\_cancel$ method. The $max\_cancel$ method fixes the logical circuit to have one leaf tree to maximize the gate cancelation. {The baseline PCOAST is further transpiled by Qiskit to solve the hardware connectivity constraint, as PCOAST is a logical circuit optimization. The $max\_cancel$ version is also further transpiled by Qiskit for the same reason. All methods except T$|$Ket$\rangle$ are optimized by Qiskit optimization level 3 (Qiskit O3).}   {It is because T$|$Ket$\rangle$ with its O2 optimization and mapping and routing passes, is better than with Qiskit O3. We denote them as T$|$Ket$\rangle$ O2. In Fig. \ref{fig:tketo2}, we show that T$|$Ket$\rangle$ + T$|$Ket$\rangle$ O2 is better than T$|$Ket$\rangle$ + Qiskit O3 in all cases, so in our experiment, we only show T$|$Ket$\rangle$ + T$|$Ket$\rangle$ O2.} For the QAOA applications, we also compare our method with 2QAN\cite{lao+:isca22}, the state-of-the-art in compiling the 2-local Hamiltonian simulation algorithm.

\paragraph{Benchmarks} We choose benchmarks of various sizes and applications from both the VQE and QAOA categories. We use the UCCSD ansatz \cite{uccsd} for VQE. We have constructed the Hamiltonians for six distinct molecules (LiH, BeH2, CH4, MgH2, LiCl, CO2) utilizing the PySCF \cite{PySCF} software package. We also have six synthetic benchmarks, denoted by UCC-10 to UCC-35, generated by randomly sampling $n^2$ blocks from the original UCCSD. For the QAOA applications, we consider two types of graphs: random graphs and regular graphs.  Table \ref{tab:benchmarks} provides a comprehensive overview of these benchmarks.



\begin{table}[ht]
\centering
\caption{{Benchmarks}}
\label{tab:benchmarks}
\begin{tabular}{|c|c|c|c|c|c|}
\hline
Type                       & Bench. & \#qubits & \#Pauli & \#CNOT & \#1Q \\ \hline
\multirow{6}{*}{Molecules} 
&    LiH    &    12      &    640     &   8064     &    4992      \\ \cline{2-6} 
&   BeH2     &    14      &   1488      &  21072   &   11712      \\ \cline{2-6} 
&  CH4      &  18        &  4240       &  73680      &  33600      \\ \cline{2-6} 
&  MgH2      &   22       &  8400       & 173264  &  66752        \\ \cline{2-6} 
&  LiCl      &  28        &  17280       & 440960 &  137600        \\ \cline{2-6} 
& CO2       &  30        &  20944       & 568656 & 166848         \\ \hline
\multirow{6}{*}{UCCSD}     
& UCC-10       &    10      &     800    &    8976    &  6400
        \\ \cline{2-6} 
& UCC-15       &     15     &   1800      &   27200     &  14400        \\ \cline{2-6} 
& UCC-20       &    20      &    3200     &   59712     &   25600       \\ \cline{2-6} 
& UCC-25       &     25     &    5000     &  117376      &  40000        \\ \cline{2-6} 
& UCC-30       &    30      &    7200     &  193984      & 57600         \\ \cline{2-6} 
& UCC-35       &    35      &   9800      &    304976    &  78400        \\ \hline
\multirow{6}{*}{QAOA}      
& Rand-16  &   16       &   25      &    50    &    57      \\ \cline{2-6} 
& Rand-18       &   18       &  31       &   62     &    67      \\ \cline{2-6} 
&  Rand-20      &    20      &    40     &   80     &      80    \\ \cline{2-6} 
& REG3-16       &    16      &   24      &   48     &    56      \\ \cline{2-6} 
&  REG3-18      &   18       &   27      &   54     &      63    \\ \cline{2-6} 
&  REG3-20      &     20     &    30     &    60    &      70    \\ \hline

\end{tabular}

\end{table}

\begin{table*}[]
\centering

\resizebox{1.0\textwidth}{!}{

\begin{tabular}{|c|c|cll|cll|cll|cll|}
\hline
\multicolumn{1}{|l|}{\multirow{2}{*}{Encoder}} & \multirow{2}{*}{Bench.} & \multicolumn{3}{c|}{Total Gate} & \multicolumn{3}{c|}{CNOT Gate} & \multicolumn{3}{c|}{Depth} & \multicolumn{3}{c|}{Duration} \\ \cline{3-14} 
\multicolumn{1}{|l|}{} &  & PH & \multicolumn{1}{c}{Tetris} & \multicolumn{1}{c|}{Improv.} & PH & \multicolumn{1}{c}{Tetris} & \multicolumn{1}{c|}{Improv.} & PH & \multicolumn{1}{c}{Tetris} & \multicolumn{1}{c|}{Improv.} & PH & \multicolumn{1}{c}{Tetris} & \multicolumn{1}{c|}{Improv.} \\ \hline
\multirow{6}{*}{Jordan-Wigner} & LiH & 9.2k & 8.8k & -4.38\% & 5.3k & 4.4k & -17.19\% & 6.2k & 5.5k & -10.98\% & 7.9M & 6.0M & -23.63\% \\ \cline{2-14} 
 & BeH2 & 25.4k & 21.1k & -17.16\% & 15.1k & 10.4k & -31.28\% & 16.0k & 12.8k & -19.88\% & 21.3M & 13.8M & -35.26\% \\ \cline{2-14} 
 & CH4 & 71.9k & 63.0k & -13.06\% & 44.9k & 31.1k & -30.78\% & 46.5k & 37.2k & -19.93\% & 60.2M & 39.9M & -33.59\% \\ \cline{2-14} 
 & MgH2 & 142.8k & 127.7k & -10.78\% & 90.3k & 63.4k & -29.79\% & 97.1k & 74.5k & -23.23\% & 130.9M & 80.6M & -38.41\% \\ \cline{2-14} 
 & LiCl & 341.3k & 273.3k & -20.20\% & 225.3k & 139.5k & -38.08\% & 245.8k & 157.9k & -35.75\% & 346.2M & 176.3M & -49.03\% \\ \cline{2-14} 
 & CO2 & 426.6k & 331.0k & -22.71\% & 284.1k & 168.5k & -40.67\% & 306.4k & 191.1k & -37.64\% & 428.2M & 209.9M & -50.94\% \\ \hline
\multirow{6}{*}{Bravyi-Kitaev} & LiH & \multicolumn{1}{l}{14.8k} & 14.1k & -4.23\% & \multicolumn{1}{l}{10.1k} & 8.5k & -16.07\% & \multicolumn{1}{l}{10.8k} & 8.7k & -19.70\% & \multicolumn{1}{l}{16.3M} & 10.8M & -33.60\% \\ \cline{2-14} 
 & BeH2 & \multicolumn{1}{l}{32.9k} & 29.3k & -11.53\% & \multicolumn{1}{l}{22.1k} & 17.3k & -21.40\% & \multicolumn{1}{l}{23.1k} & 17.8k & -23.23\% & \multicolumn{1}{l}{34.0M} & 21.5M & -36.81\% \\ \cline{2-14} 
 & CH4 & \multicolumn{1}{l}{100.7k} & 94.8k & -6.76\% & \multicolumn{1}{l}{64.2k} & 56.7k & -11.62\% & \multicolumn{1}{l}{61.5k} & 53.6k & -12.84\% & \multicolumn{1}{l}{85.0M} & 65.0M & -23.51\% \\ \cline{2-14} 
 & MgH2 & \multicolumn{1}{l}{213.5k} & 189.5k & -12.35\% & \multicolumn{1}{l}{142.2k} & 113.3k & -20.30\% & \multicolumn{1}{l}{139.7k} & 107.0k & -23.37\% & \multicolumn{1}{l}{204.2M} & 132.3M & -35.21\% \\ \cline{2-14} 
 & LiCl & \multicolumn{1}{l}{447.9k} & 393.3k & -13.09\% & \multicolumn{1}{l}{292.7k} & 232.9k & -20.40\% & \multicolumn{1}{l}{285.8k} & 215.4k & -24.65\% & \multicolumn{1}{l}{406.2M} & 265.0M & -34.75\% \\ \cline{2-14} 
 & CO2 & \multicolumn{1}{l}{561.6k} & 467.3k & -17.65\% & \multicolumn{1}{l}{385.0k} & 276.7k & -28.11\% & \multicolumn{1}{l}{381.2k} & 260.6k & -31.65\% & \multicolumn{1}{l}{567.9M} & 321.6M & -43.37\% \\ \hline
\multirow{6}{*}{Synthetic} & UCC-10 & 12.3k & 10.5k & -14.54\% & 7.6k & 5.1k & -32.89\% & 9.4k & 6.7k & -28.92\% & 12.7M & 7.2M & -43.47\% \\ \cline{2-14} 
 & UCC-15 & 28.9k & 26.3k & -9.27\% & 17.3k & 13.6k & -21.02\% & 18.9k & 16.3k & -13.88\% & 25.0M & 17.7M & -29.20\% \\ \cline{2-14} 
 & UCC-20 & 57.1k & 51.8k & -9.79\% & 36.5k & 27.9k & -23.47\% & 37.7k & 30.2k & -19.74\% & 51.4M & 33.9M & -34.13\% \\ \cline{2-14} 
 & UCC-25 & 98.8k & 87.3k & -12.11\% & 66.0k & 49.4k & -25.20\% & 64.8k & 49.6k & -23.39\% & 93.2M & 56.8M & -39.07\% \\ \cline{2-14} 
 & UCC-30 & 150.3k & 131.4k & -13.05\% & 102.3k & 76.0k & -25.70\% & 95.7k & 72.6k & -24.17\% & 136.3M & 85.0M & -37.64\% \\ \cline{2-14} 
 & UCC-35 & 212.9k & 187.8k & -12.46\% & 148.6k & 111.6k & -25.16\% & 137.3k & 101.5k & -26.37\% & 191.3M & 121.2M & -36.89\% \\ \hline
\end{tabular}
}
\caption{ The improvement (Improv.) is the percentage of gate count or depth or duration reduced by Tetris. The columns labeled 'CNOT Gate' and 'Total Gate' represent the CNOT gate count and the total gate count (SWAP decomposed into 3 CNOT). 'Total Gate' includes both 1-qubit gate and CNOT gate. PH stands for Paulihedral.}
\label{tab:comp2Paulihedral}
\end{table*}

\paragraph{Implementation Details} We implement our compiler with Qiskit library with optimization level 3 on Intel(R) Core(TM) i9-10900 CPU @ 2.80GHz - 20 cores computer. We consider Jordan-Wigner (JW) \cite{jordan+:zf28}, and Bravyi-Kitaev (BK) \cite{bravyi+:ap02} encoders.  The size of one Tetris block is set to one block of the Paulihedral block.  The Pauli-strings within a block have higher similarity than that across blocks.

\subsection{Comparison with Paulihedral, {PCOAST}, and T$|$ket$\rangle$}

\textbf{Paulihedral} \label{subsec:paulihedral} The overall comparison with Paulihedral(PH) is presented in Table \ref{tab:comp2Paulihedral}. This table showcases the experiments conducted using the Jordan-Wigner (JW) method and Bravyi-Kitaev (BK) encoders on six real molecule benchmarks {on IBM 65-qubit backend}. In general, Tetris outperforms Paulihedral. For the Jordan-Wigner encoder, Tetris has on average 30.68\% improvement in CNOT gate count; 14.76\% improvement in total gate count; and 24.62\% improvement in circuit depth.
  
Although the total gate count is not reduced as much as the 2-qubit (CNOT) gate count (this is because the 1-qubit gate count remains almost unchanged), our improvement is still important. It is because the 1-qubit gates run in parallel, but the 2-qubit gates tend to run sequentially due to the tree construction. This is reflected in the more significant reduction in circuit duration and depth. The 2-qubit gate has an order of magnitude higher error than the 1-qubit gate. Reducing the 2-qubit gate is more important. 

Bravyi-Kitaev transformer is another popular transformer that maps fermionic operators in UCCSD ansatz to Pauli operators. Compared with the Jordan-Wigner encoder, the Bravyi-Kitaev encoder tends to have slightly lower similarity among Pauli-strings. It has improvement over Paulihedral but the improvement is slightly less than that in Jordan-Wigner. Tetris has on average 17.05\% improvement in CNOT gate count; 9.72\% improvement in total gate count; and 19.19\% improvement in circuit depth. 

An important advantage of Tetris is that the improvement scales with the size of the molecules. The larger the molecule is, the more improvement Tetris has.
{All the comparison above is comparing PH+Qiskit O3 with Tetris+Qiskit O3. We also compared Tetris to PH without any Qiskit's optimization. See Fig. \ref{fig:without_qiskito3}, in both cases, with or without Qiskit optimizations, Tetris perform better than Paulihedral. Qiskit O3 improves over PH a lot because PH leaves the job of canceling gates to Qiskit O3. Tetris has its cancellation functionality, so Qiskit O3 only helps with other optimizations. We did not include PCOAST or T$|$Ket$\rangle$ into the comparison here because T$|$Ket$\rangle$'s CNOT gate count is around 2X greater than Paulihedral and Tetris so we put T$|$Ket$\rangle$'s performance into a single figure, see Fig. \ref{fig:tketo2}. And PCOAST needs Qiskit O3 to do the qubit mapping and routing, not like Paulihedral and Tetris that has finished qubit mapping and routing before Qiskit O3. So for fair comparison considering hardware connection, we can not remove the Qiskit O3 for PCOAST.}
\begin{figure}
    \centering
    \includegraphics[width=1.0\linewidth]{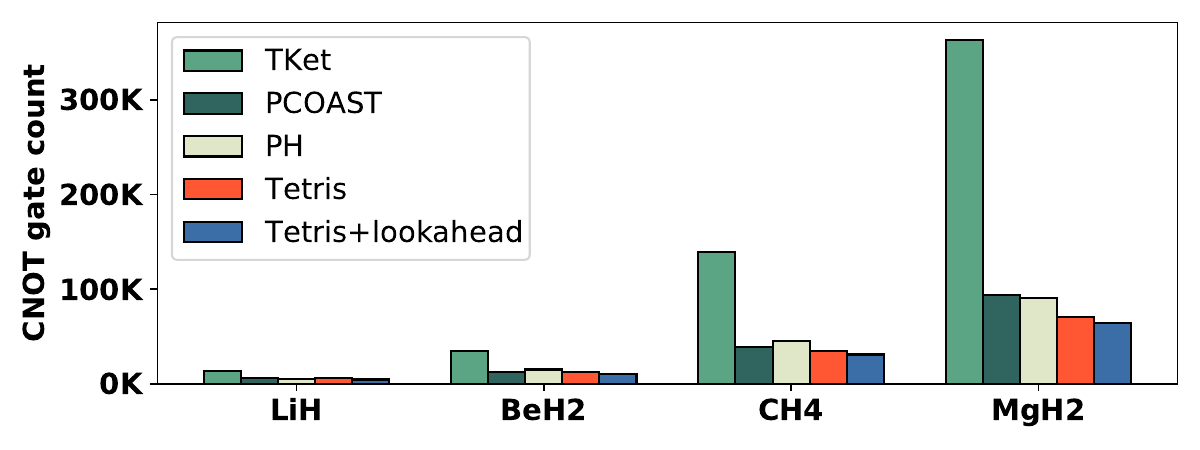}\vspace{-0.05in}
    \caption{{ Comparison: T$|$Ket$\rangle$ v.s. {PCOAST} v.s. Paulihedral v.s. Tetris v.s. Tetris+lookahead. All molecules use the JW mapper. The target coupling map is IBM heavy-hex. `Tket' is {T$|$Ket$\rangle$ + T$|$Ket$\rangle$ O2}. {`PCOAST' is intel quantum SDK O1 compilation followed by Qiskit O3.} `Tetris' is our version by replacing our scheduler with that from Paulihedral, `Tetris+lookahead' is using our version with our lookahead scheduler (K=10).}}
    \label{fig:more_baselines}
\end{figure}

\begin{figure}
    \begin{subfigure}{0.4\linewidth}
        \centering
        \includegraphics[width=\linewidth]{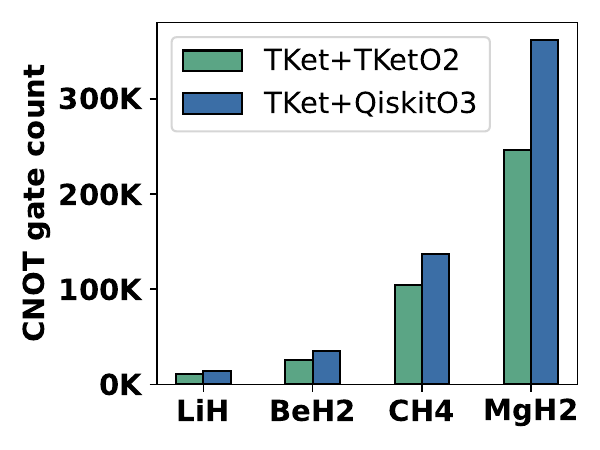}\vspace{-0.05in}
        \caption{{T$|$Ket$\rangle$ Analysis}}
        \label{fig:tketo2}
    \end{subfigure}
    \begin{subfigure}{0.6\linewidth}
        \centering
        \includegraphics[width=1\linewidth]{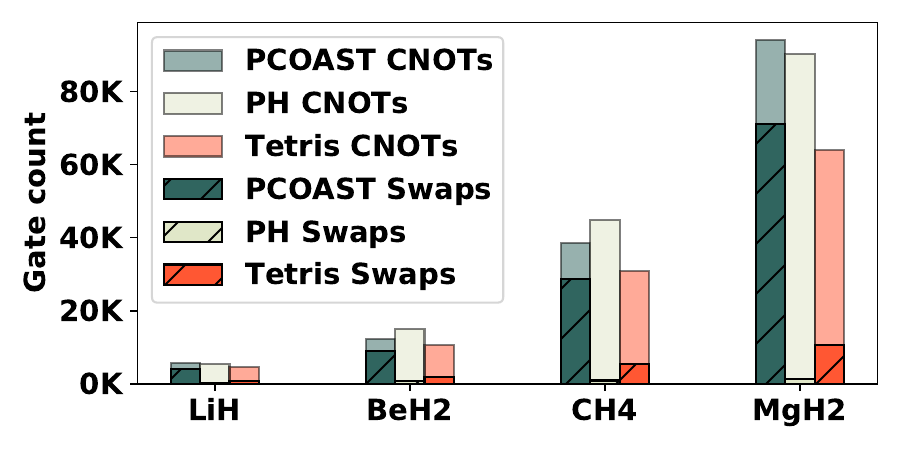}\vspace{-0.05in}
        \caption{{  Comparison between PCOAST}
        }
        \label{fig:pcoast_breakdown}
    \end{subfigure}
    \caption{{Details of T$|$Ket$\rangle$ and PCOAST. (a) Comparison between T$|$Ket$\rangle$ + T$|$Ket$\rangle$ O2 and T$|$Ket$\rangle$ + Qiskit O3. (b) Comparison between PCOAST, Paulihedral, and Tetris. ``Swaps" corresponds to the SWAP-induced CNOT count. Left bar is for PCOAST, middle bar is for Paulihedral, and right bar is for Tetris. Each bar is a breakdown of SWAP-induced CNOT and other CNOT gates. }}
    \label{fig:details_tket_pcoast}
\end{figure}

\begin{figure}
    \centering
    \includegraphics[width=1.0\linewidth]{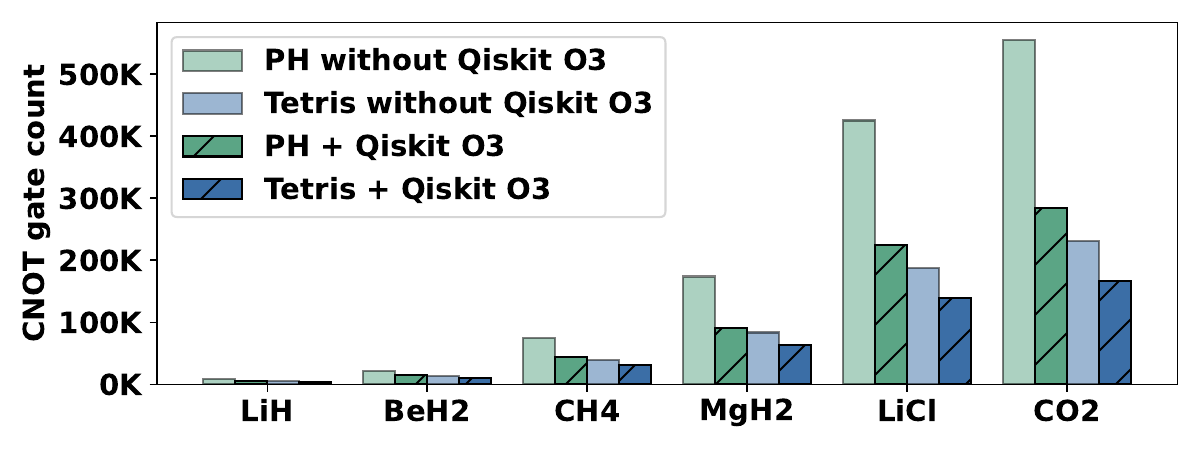}\vspace{-0.05in}
    \includegraphics[width=1.0\linewidth]{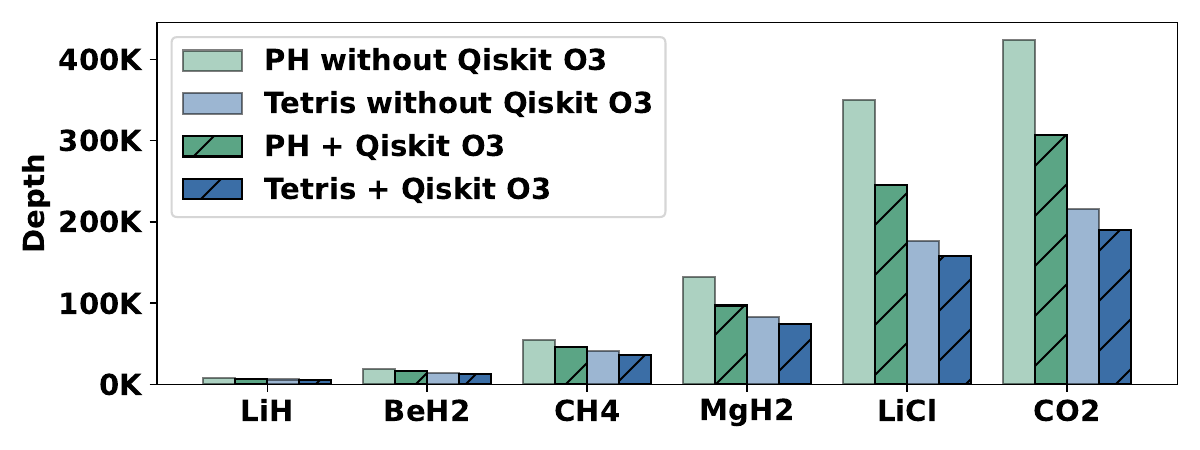}\vspace{-0.05in}
    \caption{{Comparison of PH and Tetris without Qiskit O3 and with Qiskit O3.}}
    \label{fig:without_qiskito3}
\end{figure}

\label{subsec:tket}

\textbf{{T$|$Ket$\rangle$ and PCOAST:}} {Besides Paulihedral, we compare our method with the compiler T$|$Ket$\rangle$ and PCOAST. In these comparisons, both CNOT gate count and circuit depth present a similar trend. So, we only present the result of CNOT gate count. We show the performance of T$|$Ket$\rangle$+T$|$Ket$\rangle$O2, PCOAST, Paulihedral, and Tetris with respect to different scheduling methods in Fig. \ref{fig:more_baselines}. Except for T$|$Ket$\rangle$, they are optimized with Qiskit O3. For the largest two molecules, T$|$Ket$\rangle$+O2's data is missing due to its timeout in compilation ($>$ 12 hrs). From this figure, we can see that both our methods achieve significant improvement over the baselines.}

\label{subsec:analyze_pcoast}
{In Fig. \ref{fig:pcoast_breakdown}, we show the breakdown of the swap-induced CNOT count of PCOAST. PCOAST did a very good job of minimizing the logical gate count. {PCOAST overall performs better than Paulihedral for three molecules.} PCOAST and Tetris are using different basis gate sets.  The total logical gate count generated by PCOAST using the Intel Quantum SDK's basis \{RXY, CZ\} is lower than that by Tetris using the IBM basis of \{U3, CNOT\}.  In our comparison, we have to convert 1 CZ gate into {1 CNOT gate} and some single-qubit gates. {So there's no basis gate set conversion overhead if we only consider the 2-qubit gate cost.} {While} PCOAST greatly improves the logical gate count, it doesn't consider the qubit mapping and routing problem, and the hardware-mapping-agnostic logical circuit optimization actually results in more SWAPs. {In Fig. \ref{fig:pcoast_breakdown} we can see the SWAP-induced CNOT gate is much higher than PH and Tetris.} The above factors make PCOAST not perform as good as Tetris in terms of the CNOT gate count with the consideration of hardware.}

\begin{figure}[htb]
    \centering
    \begin{subfigure}{0.5\linewidth}
        \centering
        \includegraphics[width=\linewidth]{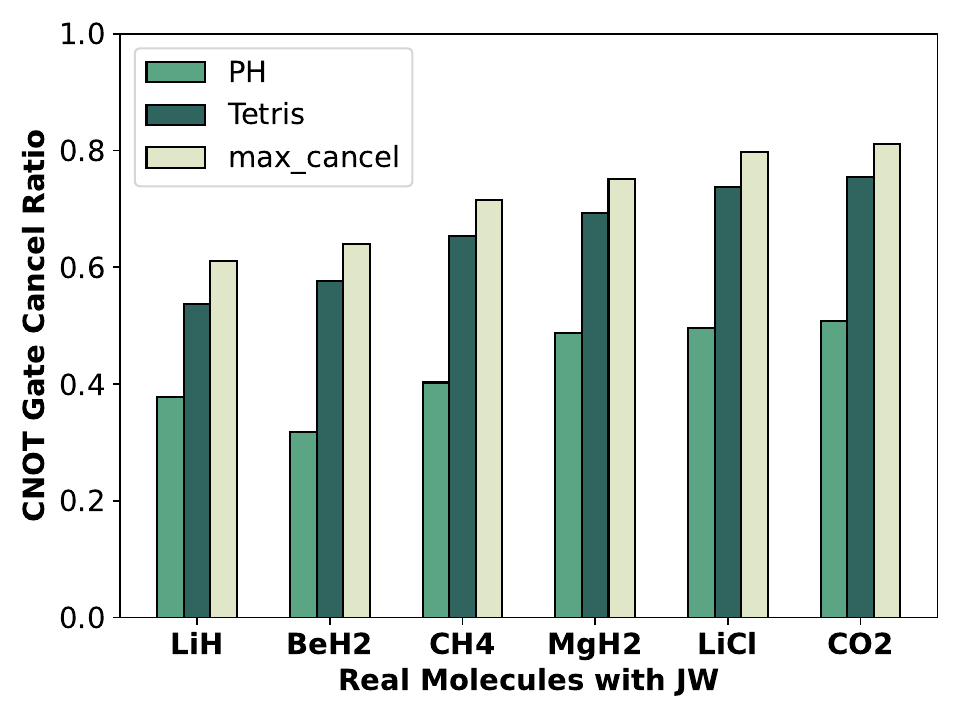}
    \end{subfigure}\hspace{-5pt}
    \begin{subfigure}{0.5\linewidth}
        \centering
        \includegraphics[width=\linewidth]{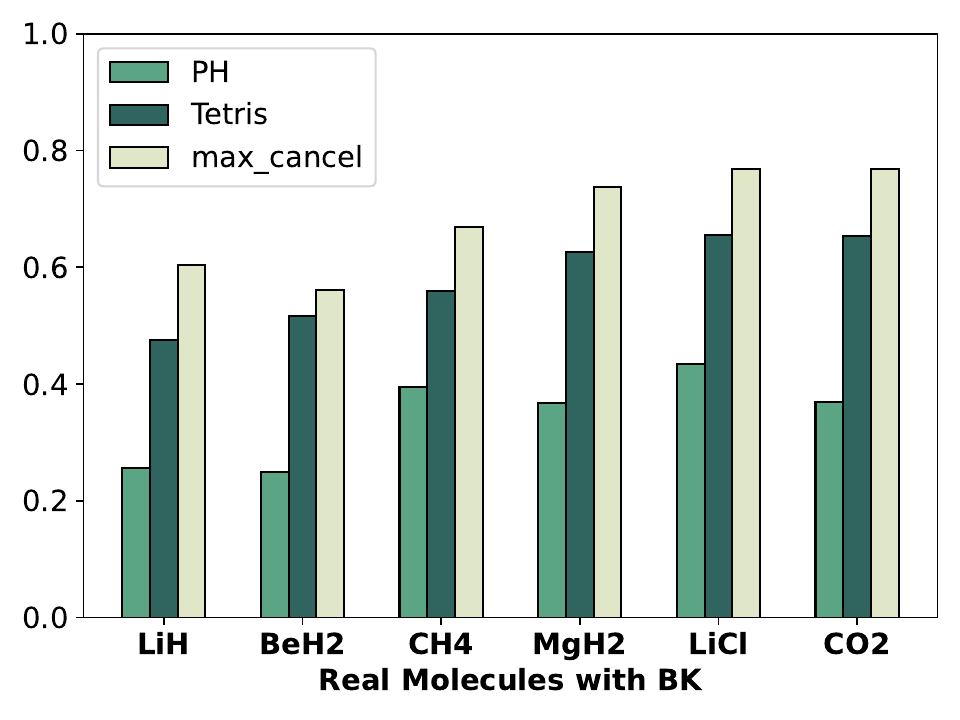}
    \end{subfigure}\vspace{-0.05in}
    \caption{Original circuit CNOT gate cancelation. PH is Paulihedral. ``max-cancel" corresponds to the logical circuit (without any SWAP insertion) and has a maximum cancellation ratio. PH has a modest cancellation ratio. Tetris in the middle.}
    \label{fig:Cancellation_Ratio_cnot}
\end{figure}



\begin{figure*}[htb]
    \centering
    \begin{subfigure}{0.3\linewidth}
        \centering
        \includegraphics[width=\linewidth]{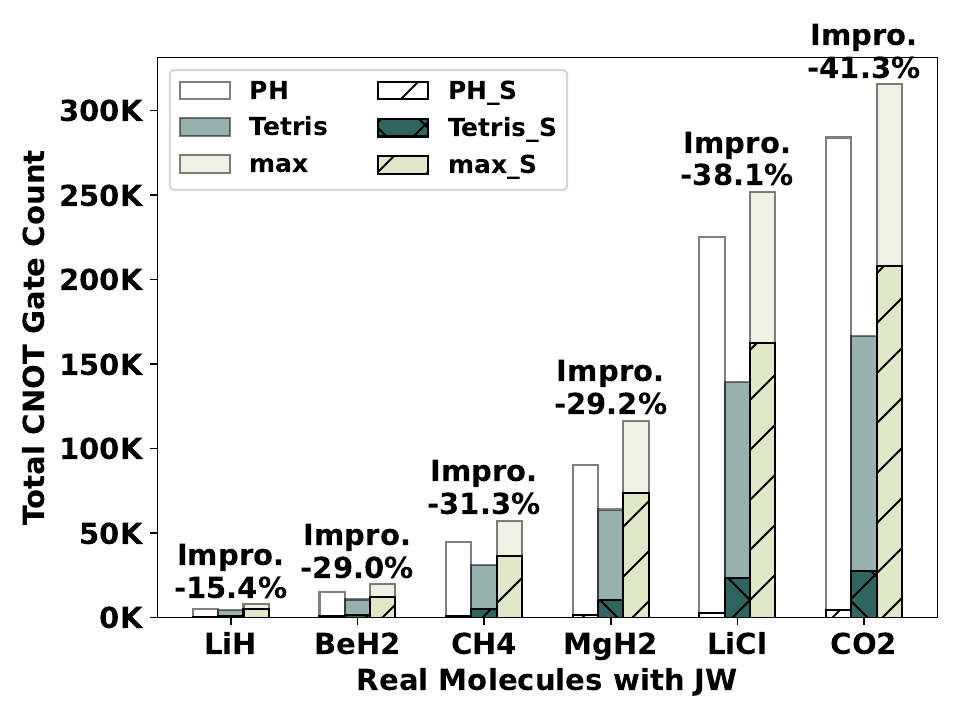}
    \end{subfigure}
    \begin{subfigure}{0.3\linewidth}
        \centering
        \includegraphics[width=\linewidth]{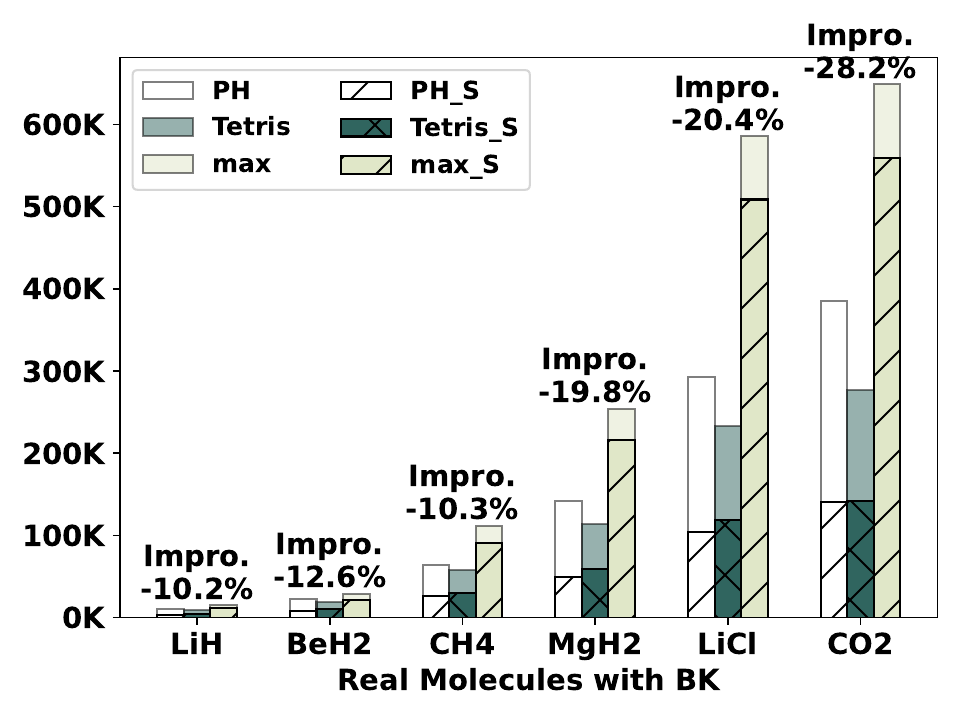}
    \end{subfigure}
    \begin{subfigure}{0.3\linewidth}
        \centering
        \includegraphics[width=\linewidth]{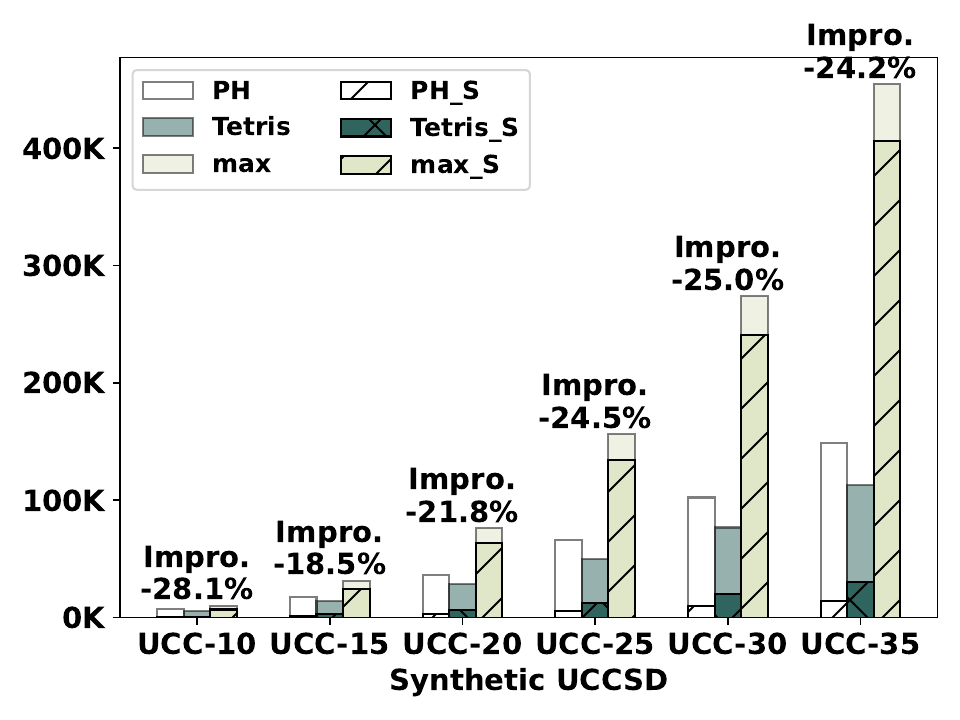}
    \end{subfigure}\vspace{-0.05in}
    \caption{Total CNOT gate breakdown analysis. Total CNOT comes from both the CNOT in the logical circuit and the CNOT in the SWAPs. The postfix ``\_S" represents the number of CNOTs from the SWAPs in each method. It shows the fraction of SWAP-induced CNOTs in each bar. ``Impro." represents the improvement of Tetris over Paulihedral in terms of gate reduction.}
    \label{fig:total_cnot_breakdown}
\end{figure*}




\subsection{Breakdown Analysis}
In this section, we analyze each individual factor in the improvement brought by Tetris, including the gate cancellation factor, and the SWAP insertion factor. PH stands for Paulihedral. The version ``max\_cancel" is an extreme case that maximizes the number of 2-qubit gate cancellations in the logical circuit. On the other hand, Paulihedral tends to prioritize SWAPs.  Tetris is the one considering both systematically.

\subsubsection{Gate Cancellation Comparison} In Fig. \ref{fig:Cancellation_Ratio_cnot}, the X-axis is the benchmark, and the Y-axis stands for the CNOT gate cancelation ratio; higher is better. From left to right, the size of the benchmark is increasing, the GCR of Tetris scales near linearly and is very close to the GCR of ``max\_cancel method". As the result shows in Fig. \ref{fig:Cancellation_Ratio_cnot}, Tetris performs well and achieves a middle ground of the cancelled logical CNOT gates between the ``max\_cancel" method and Paulihedral.

Compared with Pauliheral, the main advantage of Tetris is that Tetris guarantees a certain level of CNOT gate cancellation not be limited by the hardware connectivity constraint.



\subsubsection{SWAP Cost Analysis}  We show the total CNOT gate count comparison in Fig. \ref{fig:total_cnot_breakdown} with the breakdown. The X-axis is still the benchmarks and the Y-axis stands for the total CNOT gates. Each bar in this figure is split into two parts: the canceled logical CNOT gates, and the CNOT counts from the inserted SWAPs for solving the connectivity constraint. Paulihedral has the minimum number of CNOT gates from SWAP insertion for each benchmark, as expected. Tetris inserts more SWAP gates than Paulihedral but introduces much fewer SWAPs than the ``max\_cancel method". In benchmarks using the Jordan-Wigner encoder, although Tetris introduces more SWAPs, the fraction of SWAP-induced CNOT is small.  In benchmarks using the Bravyi-Kitaev encoder, the count of inserted SWAP gates by Tetris is close to the SWAP cost by Paulihedral. In both cases, Tetris has a better overall CNOT gate count than Paulihedral.

\vspace{-2pt}
\subsection{\fixit{Sensitivity Analysis}}
\label{sec:sensitivityanalysis}
\vspace{-4pt}

\begin{figure*}
    \centering
    \begin{subfigure}{0.16\linewidth}
        \centering
        \caption{LiH}
        \includegraphics[width=\linewidth]{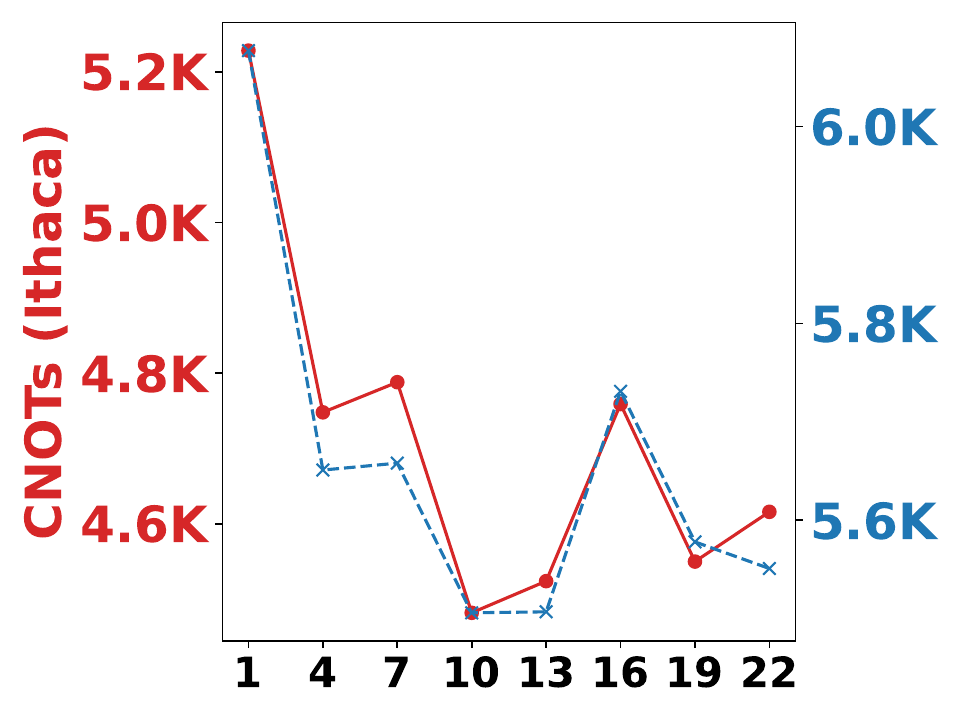}
    \end{subfigure}
    \begin{subfigure}{0.16\linewidth}
        \centering
        \caption{BeH2}
        \includegraphics[width=\linewidth]{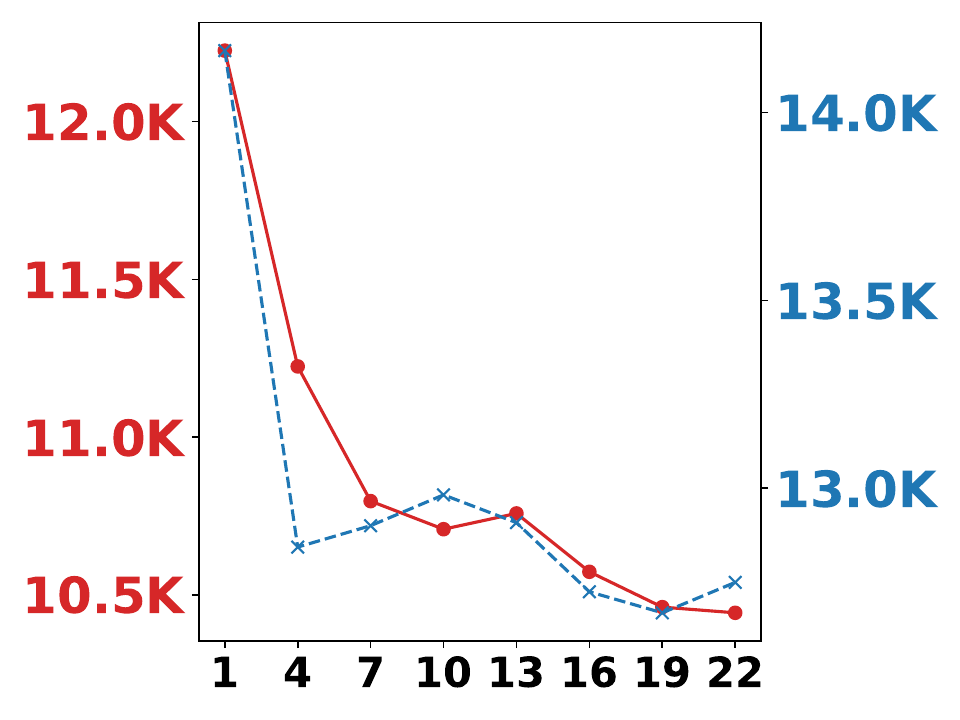}
    \end{subfigure}
    \begin{subfigure}{0.16\linewidth}
        \centering
        \caption{CH4}
        \includegraphics[width=\linewidth]{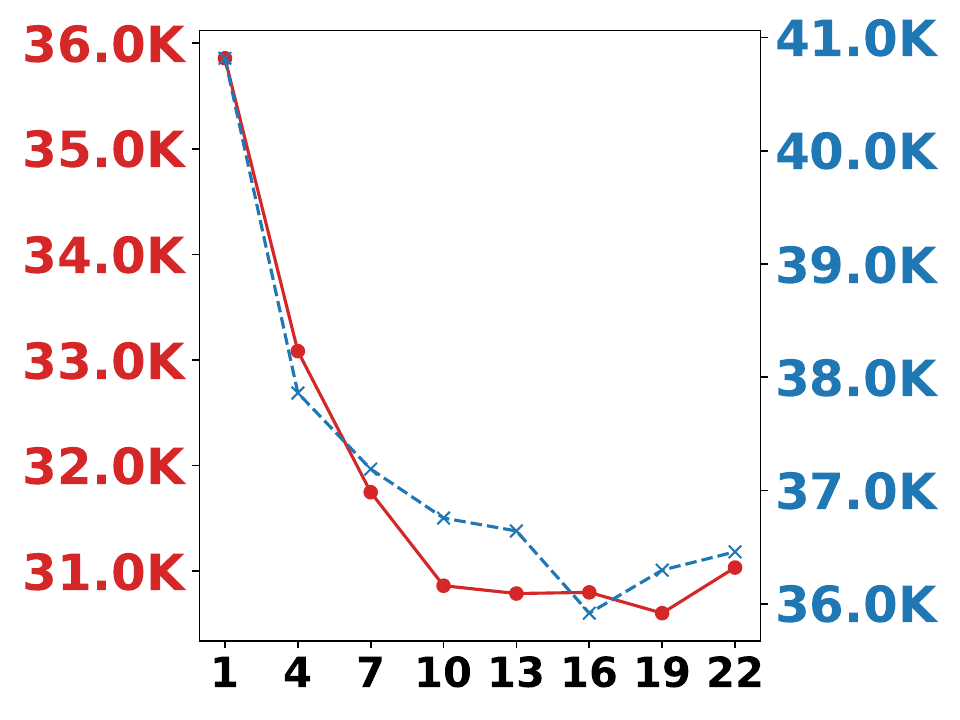}
    \end{subfigure}
    \begin{subfigure}{0.16\linewidth}
        \centering
        \caption{MgH2}
        \includegraphics[width=\linewidth]{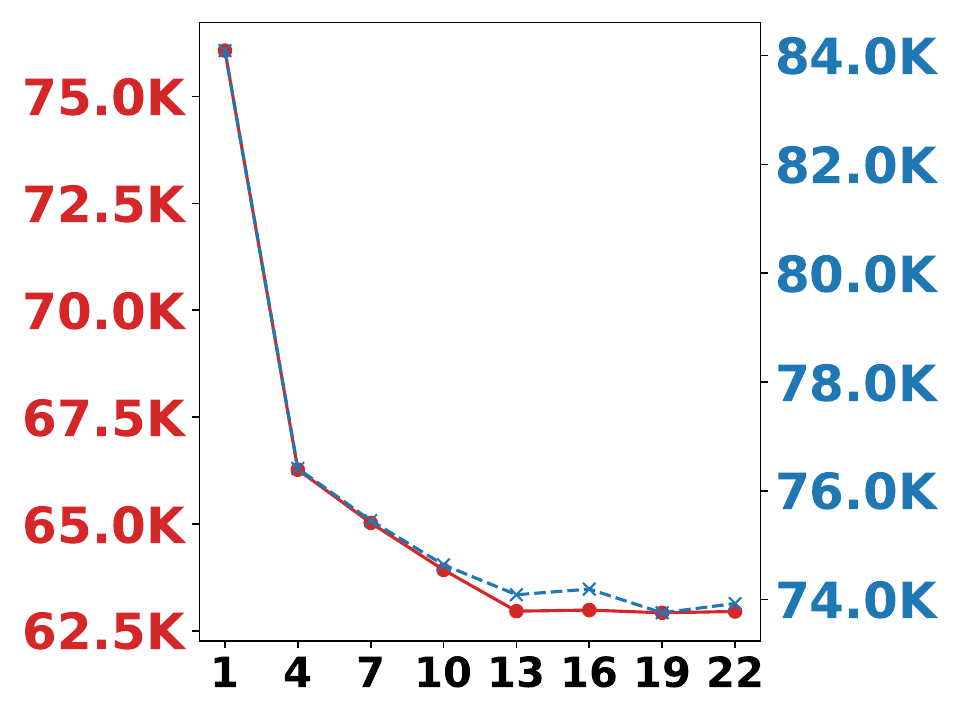}
    \end{subfigure}
    \begin{subfigure}{0.16\linewidth}
        \centering
        \caption{LiCl}
        \includegraphics[width=\linewidth]{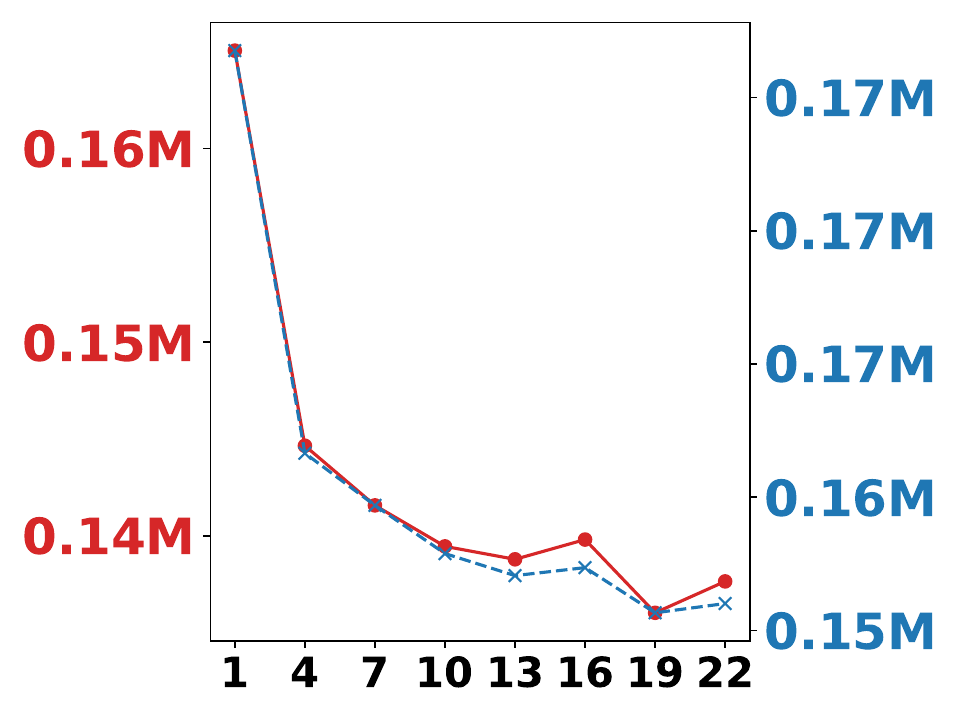}
    \end{subfigure}
    \begin{subfigure}{0.16\linewidth}
        \centering
        \caption{CO2}
        \includegraphics[width=\linewidth]{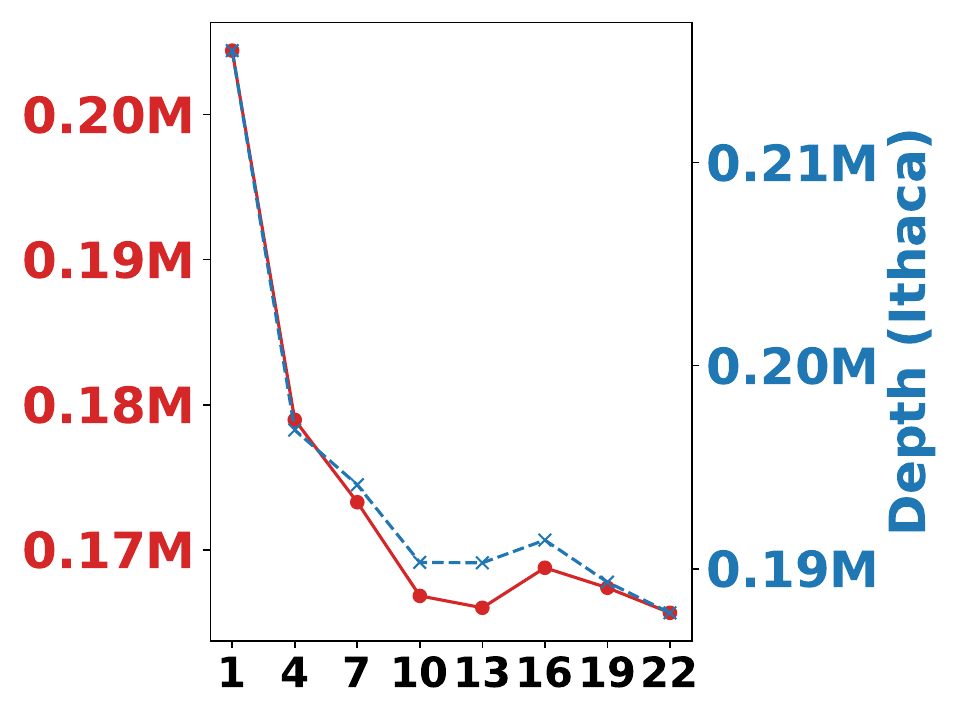}
    \end{subfigure}
    \caption{\fixit{The X-axis stands for the lookahead size K, and the Y-axis stands for total CNOT gate counts and circuit depth. The six molecules are using JW mapper.}}
    \label{fig:lookaheadK}
\end{figure*}

\fixit{\textbf{Lookahead K:} We demonstrate experiment results in IBM heavy-hex architecture by varying the value of $K$ from 1 to 22, in Fig. \ref{fig:lookaheadK}.  
The CNOT gate count is high when K = 1; almost no lookahead is enabled. By increasing the lookahead, the CNOT gate count drops quickly. It demonstrates our lookahead strategy is effective.}   

\fixit{Increasing the lookahead size does not always guarantee improvement because it is a local greedy block scheduling method. For experiments with K $>$ 10, the total CNOT gate count in the compiled circuit remains relatively stable. Therefore, we set $K = 10$ for all following experiments.}

\fixit{\textbf{SWAP Weight Analysis:}
We analyze how the cost function mentioned in Sec. \ref{sec:syntech} affects the tradeoff between SWAP reduction and logical CNOT canceling in different architectures. We vary the weight $w$ from 0.1 to 100. The compiler chooses between favoring fewer SWAP gates or more logical CNOT canceling by varying the SWAP weight. If the weight $w$ is small, the compiler favors logical gate canceling by connecting leaf qubits in a way that maximizes canceling. If the weight $w$ is large, the compiler favors connecting a qubit to a nearby mapped qubit, regardless of whether it leads to a better-synthesized tree for gate canceling. }

\begin{figure*}
    \centering
    \begin{subfigure}{0.16\linewidth}
        \centering
        \caption{BeH2}
        \includegraphics[width=\linewidth]{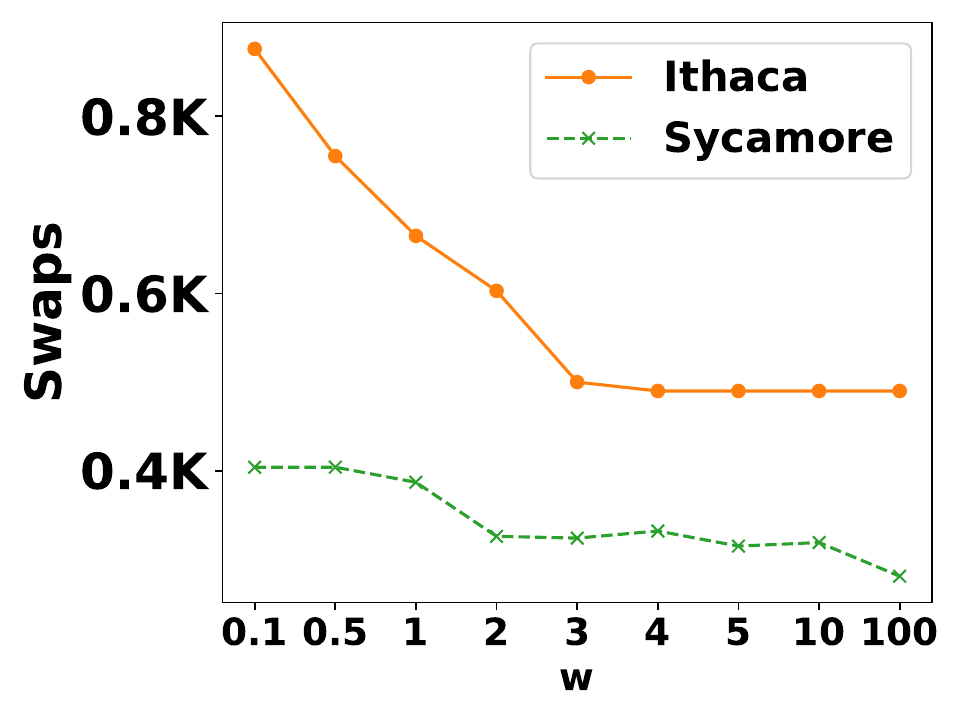}
        \label{mole1_swap_w}
    \end{subfigure}
    \begin{subfigure}{0.16\linewidth}
        \centering
        \caption{MgH2}
        \includegraphics[width=\linewidth]{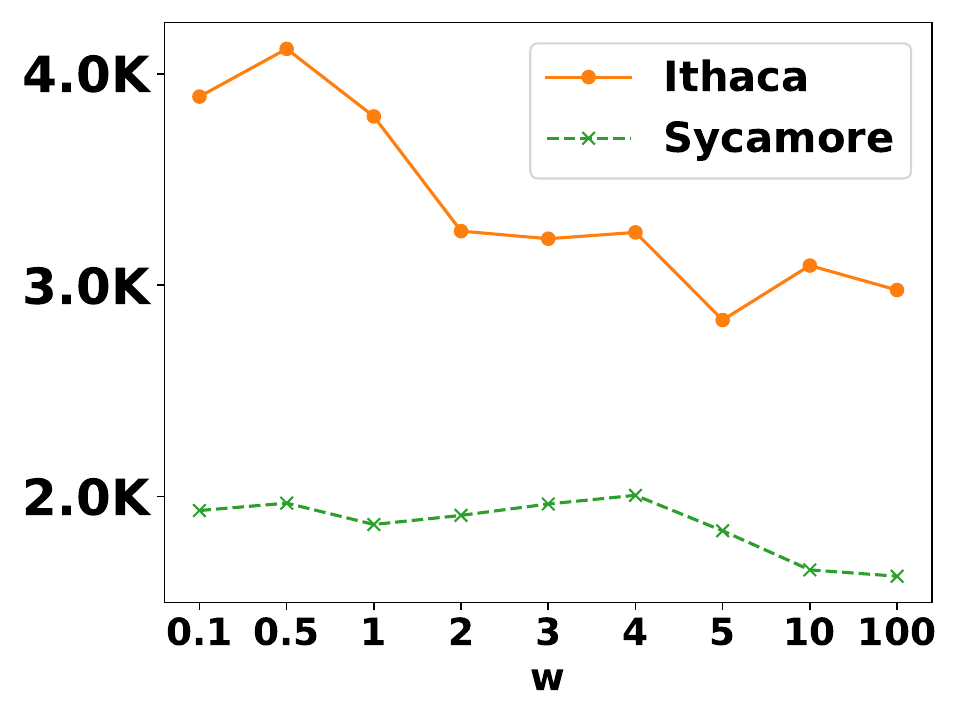}
        \label{mole3_swap_w}
    \end{subfigure}
    \begin{subfigure}{0.16\linewidth}
        \centering
        \caption{CO2}
        \includegraphics[width=\linewidth]{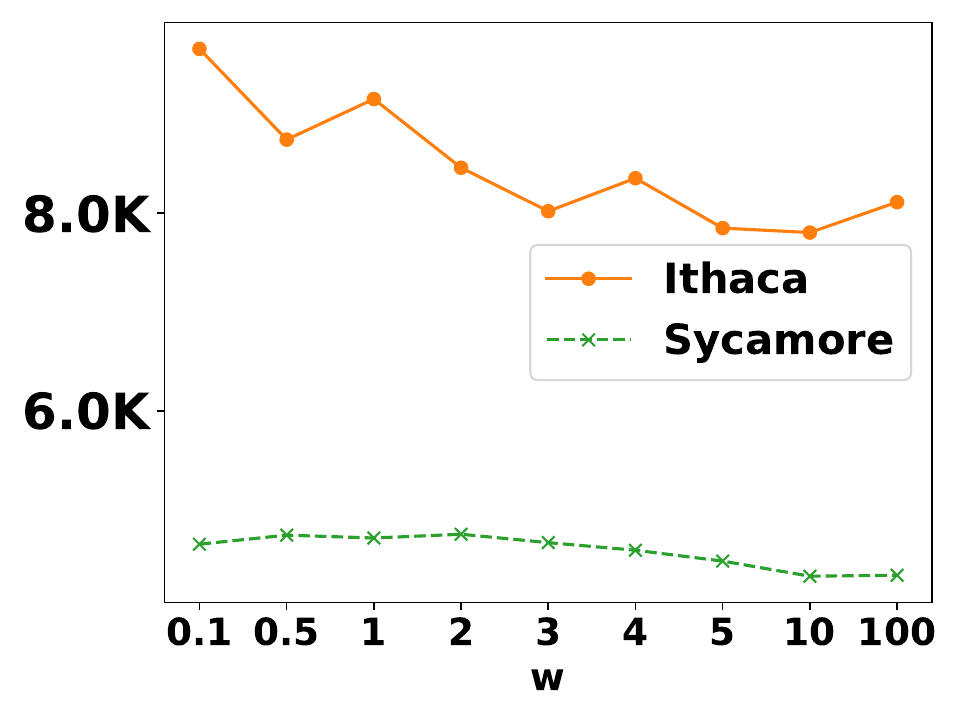}
        \label{mole5_swap_w}
    \end{subfigure}
    \begin{subfigure}{0.16\linewidth}
        \centering
        \caption{BeH2}
        \includegraphics[width=\linewidth]{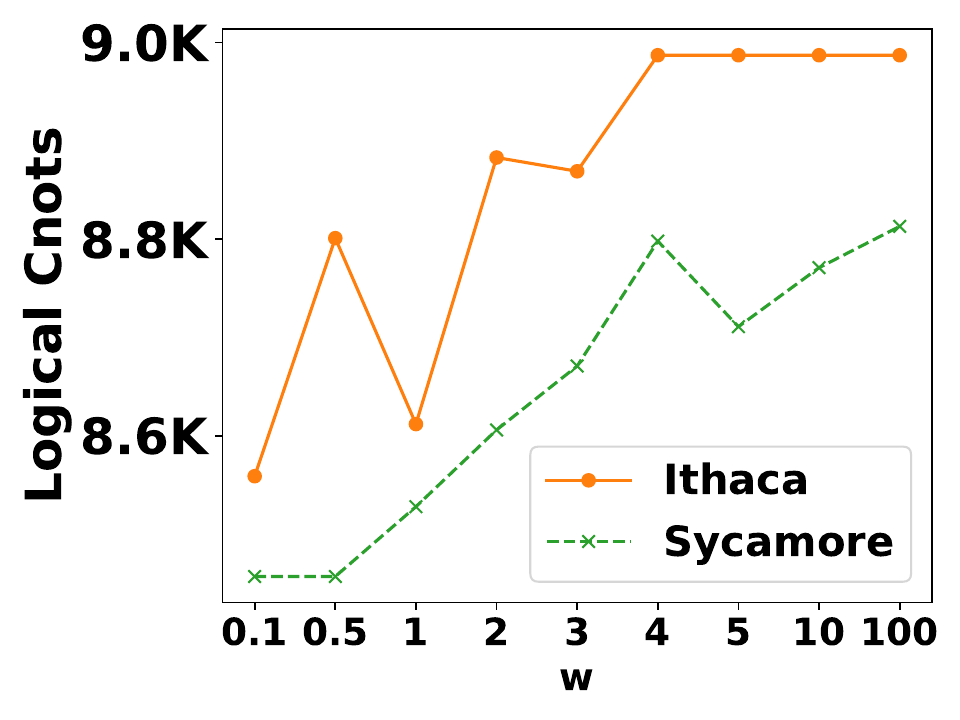}
        \label{mole1_logical_cnots_w}
    \end{subfigure}
    \begin{subfigure}{0.16\linewidth}
        \centering
        \caption{MgH2}
        \includegraphics[width=\linewidth]{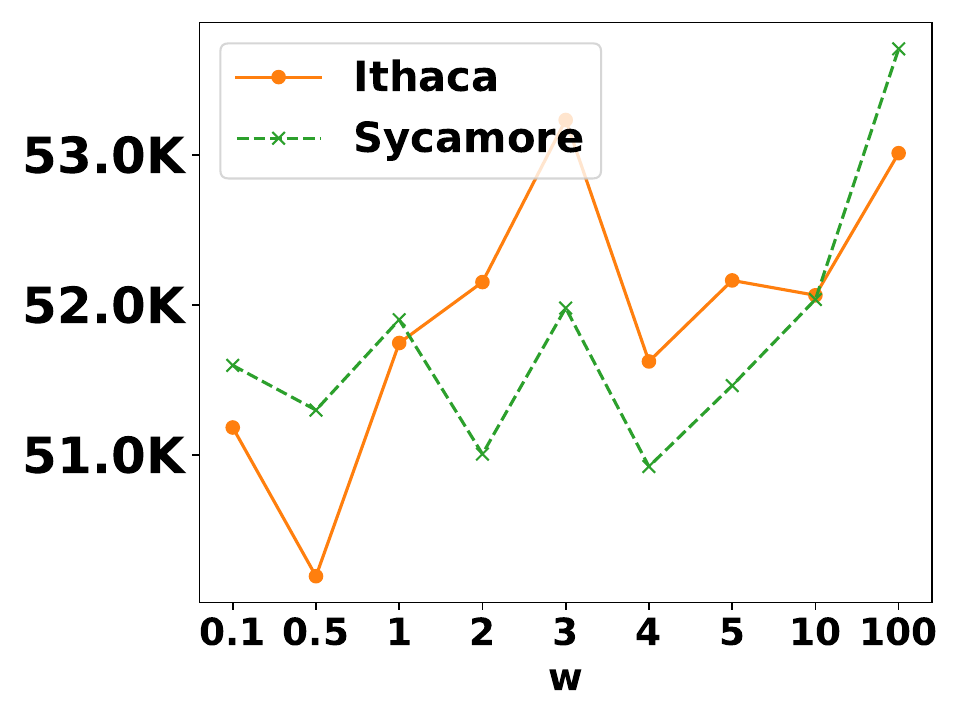}
        \label{mole3_logical_cnots_w}
    \end{subfigure}
    \begin{subfigure}{0.16\linewidth}
        \centering
        \caption{CO2}
        \includegraphics[width=\linewidth]{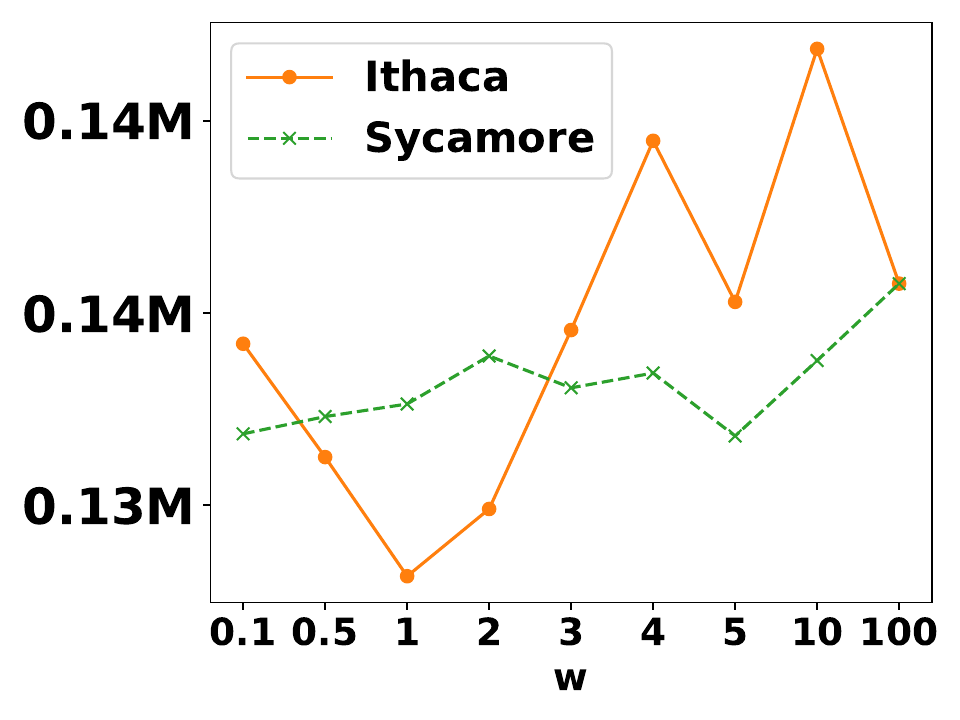}
        \label{mole5_logical_cnots_w}
    \end{subfigure}
    \vspace{-15pt}
    \caption{\fixit{Swap Weight Analysis. The X-axis stands for the swap weight $w$. (a)-(c) show the trend of SWAP count. (d)-(e) show the trend of logical  CNOT gate count.}}
    \label{fig:swapWeigth}
\end{figure*}

\fixit{In Fig. \ref{fig:swapWeigth} we show such tradeoff. When the weight $w$ increases, both architectures' SWAP gate count decreases, but the logical CNOT gate count increases (although fluctuating). The SWAP count remains low and relatively stable for each benchmark in Google Sycamore. This is because Google Sycamore architecture has higher connectivity and allows multiple shortest paths, among which we can choose the best for logical gate canceling. }

\subsection{Effects on A Different Architecture}
\label{sec:differentarchitecture}
\begin{figure}[htb]
    \centering
    \begin{subfigure}{0.5\linewidth}
        \centering
        \includegraphics[width=\linewidth]{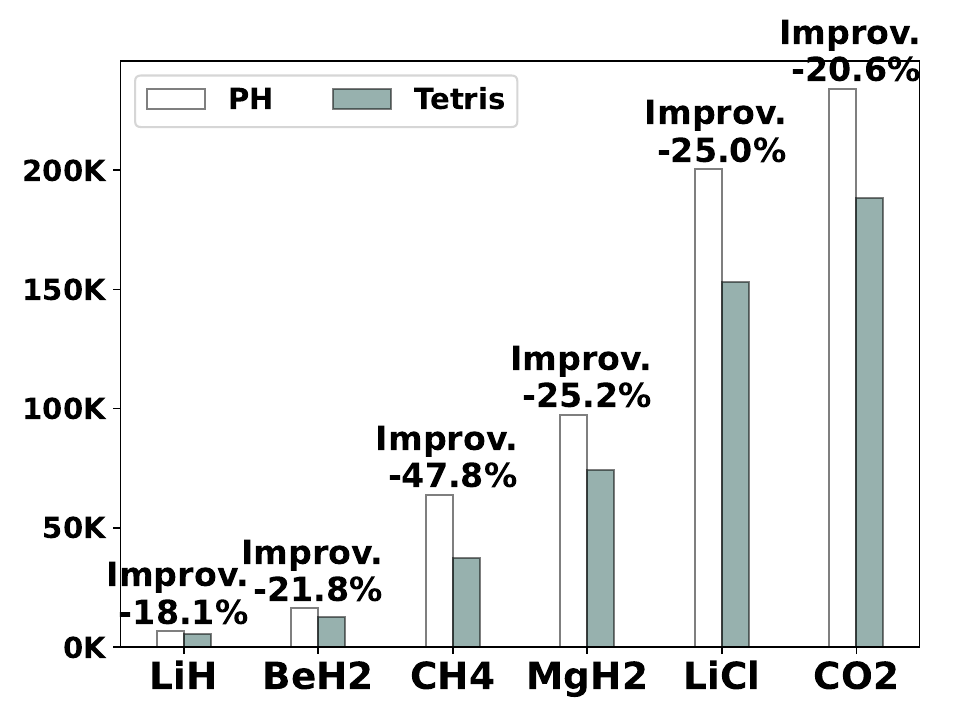}
        \caption{Depth Comparison}
    \end{subfigure}
    \hspace{-8pt}
    \begin{subfigure}{0.5\linewidth}
        \centering
        \includegraphics[width=\linewidth]{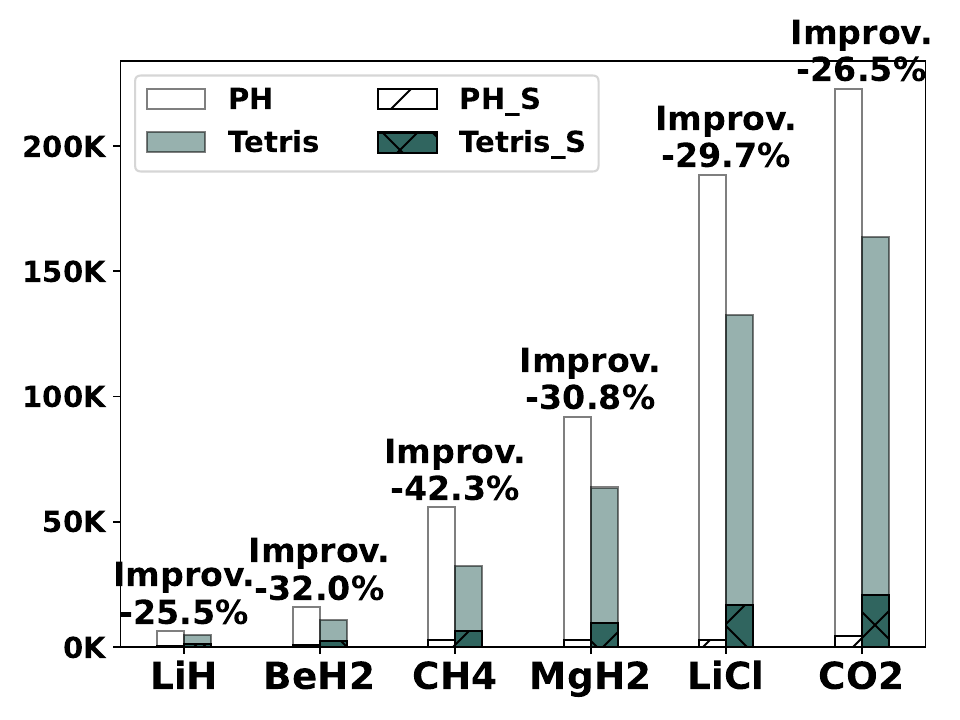}
        \caption{Total CNOT  Comparison}
    \end{subfigure}
    \caption{Comparison between Tetris and Paulihedral on Sycamore architecture for the benchmarks using Jorden-wigner encoder. The postfix ``\_S" corresponds to the SWAP-induced CNOT count. ``improv." is the relative reduction in  depth or gate count by Tetris. \vspace{-4pt}}
    \label{fig:sycamore_result}
\end{figure}
This section analyzes how the hardware connectivity affects the circuit compilation result. We use the Google Sycamore backend with six molecule benchmarks. Google Sycamore architecture has denser connectivity. \fixit{This could help both compilers reduce the SWAP cost. However, it also increases the chance of gate cancelation for Paulihedral. But overall, our approach outperforms Paulihedral in both depth and total CNOT count. }

\vspace{-2pt}
\subsection{{QAOA Benchmarks}}
\vspace{-4pt}
In this section, we discuss the performance of bridging optimization on QAOA applications. We compare our method with Paulihedral and 2QAN. As the result shown in Fig. \ref{fig:qaoacompare}, the X-axis stands for the benchmarks and the Y-axis stands for the circuit depth and total CNOT count result, normalized to Paulihedral. The smaller, the better. For the random graphs, we set the graph density to 0.1. For regular graphs, we set the degree of each node  to 3.  For each type of graph, we generate five cases randomly utilizing the Python networkx library and show the result by taking the average. Both 2QAN and Tetris have significant improvements over Paulihedral.  Tetris has even better results than 2QAN. Tetris has on average 15.3\% and 66.5\%  depth reduction over 2QAN and Paulihedral, respectively; and 20.2\% and 60.6\%  gate count reduction over 2QAN and Paulihedral, respectively.

\subsection{Fidelity Analysis}
\label{sec:fidelityanalysis}
\vspace{-4pt}
\begin{figure}[htb]
    \centering

    \begin{subfigure}{0.45\linewidth}
        \centering
        \includegraphics[width=\textwidth]{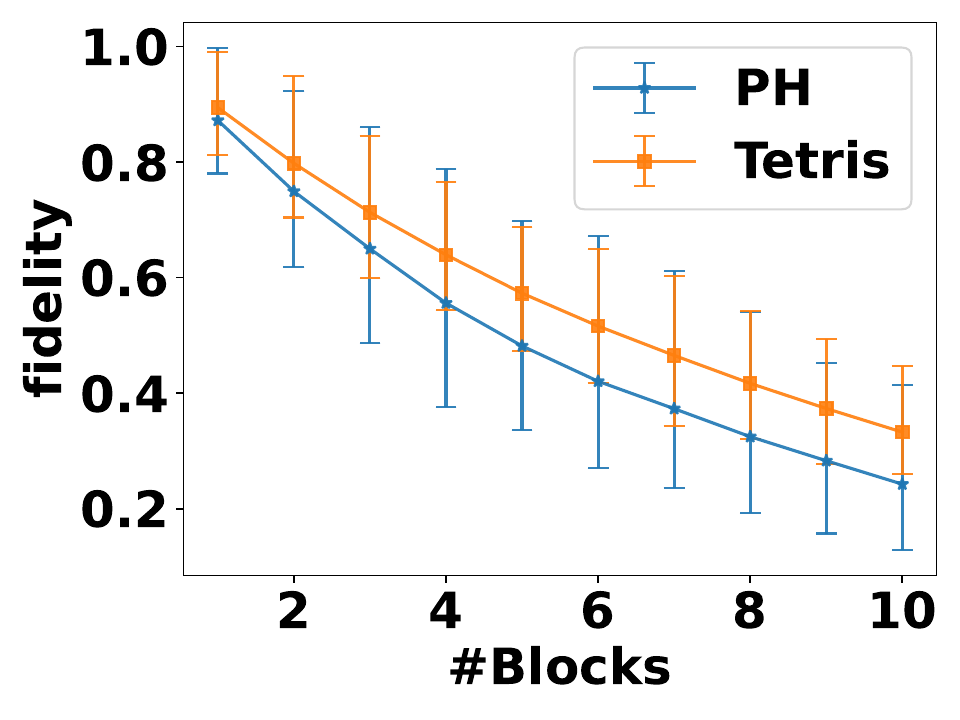}\vspace{-0.05in}
        \caption{LiH}
    \end{subfigure}
    \begin{subfigure}{0.45\linewidth}
        \centering
        \includegraphics[width=\textwidth]{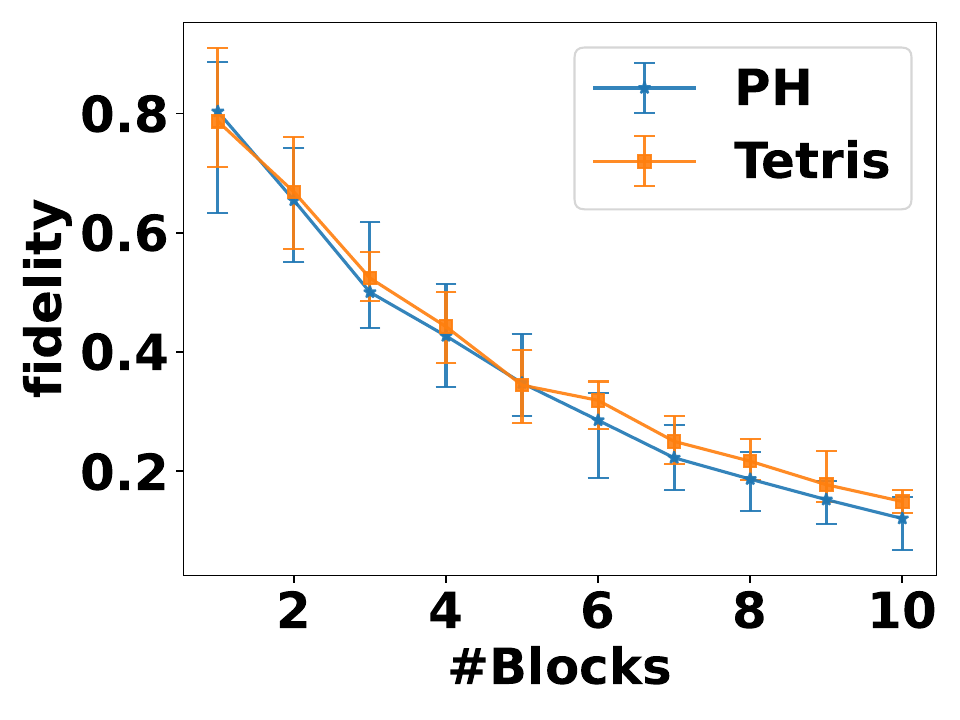}\vspace{-0.05in}
        \caption{CO2}
    \end{subfigure}\vspace{-0.05in}
    \caption{{Noise simulation results of two molecules with different numbers of Pauli blocks (the higher the better). For LiH we sampled 100 times. We use box plot to describe the min, max, and average. The points on the curve are the average of the samples for each configuration. For the biggest molecule CO2, we sampled 10 times.}}
    \label{fig:fidelity}
\end{figure}

\begin{figure}[htb]
    \centering
    \begin{subfigure}{0.4\linewidth}
        \centering
        \includegraphics[width=\linewidth]{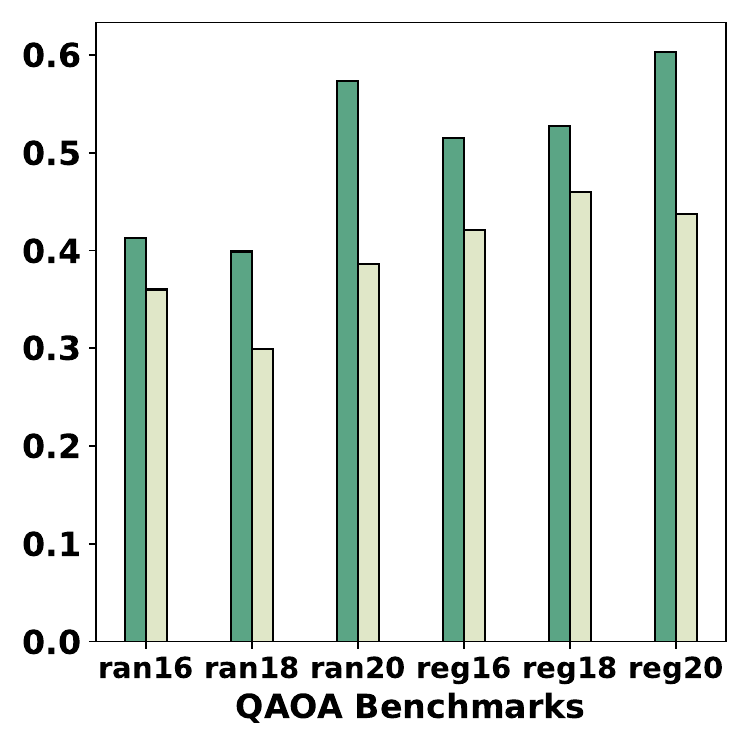}
        \caption{Gate Count Comparison}
    \end{subfigure}
    \hspace{-8pt}
    \begin{subfigure}{0.4\linewidth}
        \centering
        \includegraphics[width=\linewidth]{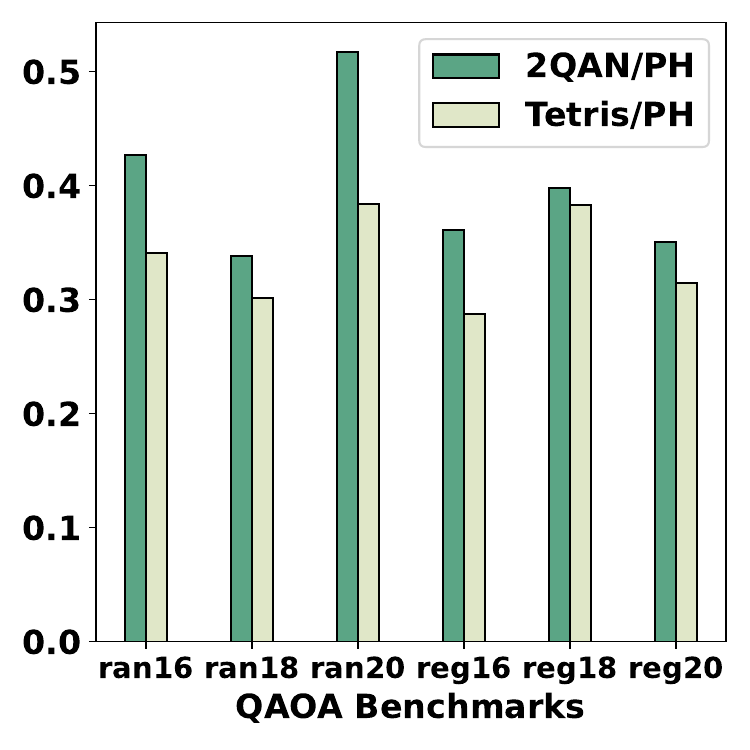}
        \caption{Depth Comparison}
    \end{subfigure}\vspace{-0.05in}
    \caption{Comparison for the QAOA benchmarks. The random graph is represented as 'ran' with different qubit numbers and 'reg' represents the regular graph.  Normalized to \fixit{Paulihedral}. }\vspace{-0.1in}
    \label{fig:qaoacompare}
\end{figure}


\fixit{
\textbf{Fidelity on Noise Simulator:}
We use the Qiskit Aer simulator with a noise model to measure circuit fidelity.
Our definition of fidelity is the same as that used in IBM random benchmarking {\cite{magesan+:prl11}}. By running a circuit followed by the inverse of the circuit on the noise simulator, we measure the probability that the state collapsed onto $|00\dots0\rangle$, regarded as fidelity in Fig~\ref{fig:fidelity}. Ideally, if there is no noise, the probability of measuring all 0's is 100\%. }

\fixit{
{Since the benchmarks we target in this work have gate counts that are beyond what NISQ computers can execute with any accuracy.
So, we make two simplifying assumptions to demonstrate the fidelity improvement with our compilation method relative to the baseline.
}
{The first simplification is we include in the circuit only 1 to 10 blocks {by random sampling} from the full set.}
{The second simplification is we choose as the noise model a depolarizing channel with parameter $10^{-3}$ for all CNOT gates and depolarizing channel with parameter $10^{-4}$ for all single qubit gates. Such a noise model represents higher fidelity than current NISQ machines. But it represents the error rates expected in future generations of quantum hardware. Using such a model, we show that our compilation method shows consistent benefit compared with Paulihedral in Fig~\ref{fig:fidelity}.}}
\vspace{-2pt}
\subsection{Scalability Analysis }
\begin{figure}[htb]
    \centering
\includegraphics[width=0.4\textwidth]{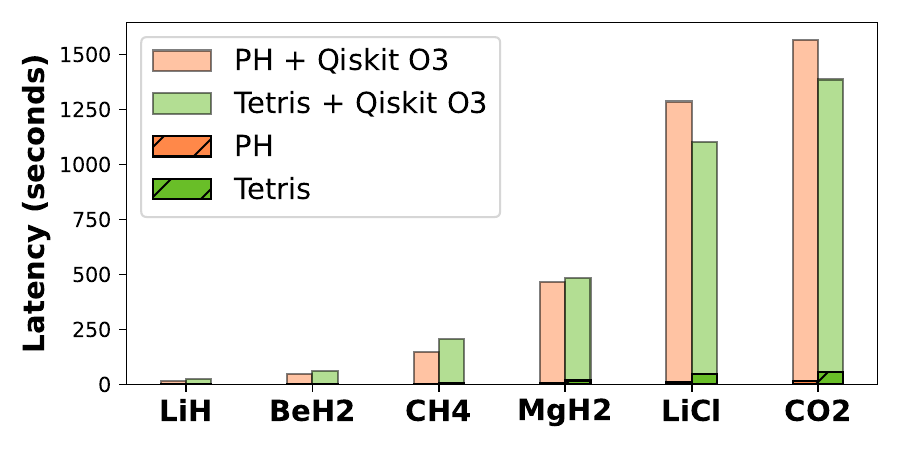}\vspace{-0.1in}
    \caption{\fixit{Scalability Analysis.}}
    \label{fig:scalability_analysis}

\end{figure}

\vspace{-4pt}
We show the scalability analysis in Fig. \ref{fig:scalability_analysis}, where the x-axis is the benchmark and the y-axis is the compilation time in seconds. The compilation time includes the compilation time from the proposed method and the subsequent optimization time consumed by Qiskit. In this figure, the Tetris compiler is faster than Paulihedral. 
This is due to Tetris having the smallest total gate count. Since Qiskit "O3" optimization further reduces gate count from the circuit compiled by Tetris and Paulihedral, the Qiskit overhead is also significantly reduced for Tetris compared with Paulihedral.
\fixit{ In Fig. \ref{fig:scalability_analysis}, we can see our proposed method Tetris has longer compilation times than PH, but the overall latency considering Qiskit "O3" is lower than PH when the molecules scale.}

\section{Related Work}

For VQE, several studies aim to optimize the circuit size from both the initial Hamiltonian construction stage \cite{peruzzo+:nature14, kandala+:nature17, lee+:jctc19}, the fermionic to spin operator encoding stage \cite{bravyi+:ap02, jordan+:zf28, seeley+:jcp12}, the circuit synthesis stage \cite{li+:asplos22, dangwal+:varsaw2023, wang:hpca2022quantumnas}, the hardware mapping stage \cite{li+:isca21, li+:asplos22,hua+:arxiv_crosstalk},  measurement grouping  \cite{dangwal+:varsaw2023, paykin2023pcoast, tqe2020:Gokhale+}, and pulse level optimization \cite{liang2023spacepulse, liang2024napa,chen+:hpca23_pulse}. Some other studies focus on the compilation of VQA circuits on different quantum platforms, such as neutral atom array \cite{wang:dac2023q,wang:isca2024fpqac}. Our paper focuses on the circuit synthesis and the hardware mapping stage. In our setting, we use the most commonly used unitary coupled cluster (UCC) ansatz \cite{peruzzo+:nature14} with single and double (SD) electron excitation operators. We also use the popular Jordan Wigner and Bravyi Kitaev \cite{bravyi+:ap02, jordan+:zf28} encoders to transform fermionic operators into spin operators. 

Li \etal \cite{li+:isca21} develops a software-hardware co-design framework for circuit synthesis and hardware mapping of VQE. It performs circuit synthesis on a proposed $X-$tree architecture. It also prunes the number of Pauli-strings with respect to their importance. Paulihedral \cite{li+:asplos22} for the first time defines the syntax and semantics of the Pauli-string IR. However, their approach focuses on single-qubit canceling, not necessarily two-qubit canceling. Moreover, it has an emphasis on SWAP reduction. It may have overlooked the potential of two-qubit gate canceling while focusing on SWAP insertion and single-qubit gate cancellation.

\label{sec:modifiedrelatedwork}
\fixit{Wang \etal in \cite{wang2023}  propose multiple high-level circuit optimizations to reduce the CNOT gate counts. It reduces the CNOT by using a new bosonic and hybrid encoding. It also performs logical CNOT gate canceling by shaping each Pauli-tree in a star-like structure and by circuit commuting. The difference is that their work is only for logical circuits rather than physical circuits that need SWAP gates. Their tree structure is fixed, while our tree structure is flexible. Yen \etal \cite{npj2022:yen+} and Gokale \etal \cite{tqe2020:Gokhale+} reduce the number of measurements from $O(N^4)$ to $O(N^3)$. As our work focuses on ansatz optimization, our work is complementary to theirs. PCOAST \cite{paykin2023pcoast} aims to reduce circuit complexity by applying graph-based representation for circuit optimization, leveraging commutative properties of Pauli operators. However, it focuses on the logical circuit optimization. 
}
\eddy{BQSKit\cite{bqskit} is a circuit synthesis tool that performs global circuit optimization on the logical circuit level to reduce the depth of quantum circuits. }

\section{Conclusion}

We introduce Tetris, a novel compilation framework designed for VQA computation kernels. Tetris recognizes an overlooked opportunity for reducing the two-qubit gate count due to the similarity of Pauli strings. Tetris also presents a unique Intermediate Representation (IR) that simultaneously addresses two-qubit gate cancellation and SWAP insertion for superconducting hardware. 

\section*{acknowledgements}
We thank the anonymous reviewers for
their constructive and helpful feedback. 
This work is supported in part by grants from the Rutgers Research Council, NSF-2129872, NSF-1818914 and 2325080 (with a subcontract to NC State University from Duke University), and 2120757 (with a subcontract to NC State University from the University of Maryland). Any opinions, findings, conclusions, or recommendations expressed in this material are those of the authors and do not necessarily reflect the views of our sponsors.

%
%
%
%
%


\appendix
\section*{Artifact Evaluation}

\subsection{Abstract}

This artifact provides the source code for the Tetris compiler and other necessary code to run the baselines in our evaluation. It also provides the scripts to reproduce the key results (Table \ref{tab:comp2Paulihedral}, Fig. \ref{fig:more_baselines}, and Fig. \ref{fig:fidelity}).

All the hardware and software you need is a computer with a Linux system. We will compare the circuit duration and fidelity using qiskit's noise simulators. The configurations of the noise simulators are all in the artifact. All the experiments will run locally on your computer and no internet connection is needed after installation.

\subsection{Artifact check-list (meta-information)}


{\small
\begin{itemize}
  \item {\bf Algorithm:} Tetris has 2 algorithms:
  
  - Tetris: circuit synthesis w.r.t. hardware (Sec. \ref{sec:syntech}) is in the function `try\_block' in `core/utils/synthesis\_lookahead.py'
  
  - Lookahead block scheduling (Sec. \ref{sec:blockscheduling}) is in the function `synthesis\_lookahead' in `core/utils/synthesis\_lookahead.py'
  
  \item {\bf Program: } \href{https://www.ibm.com/quantum/qiskit}{Qiskit}, \href{https://www.quantinuum.com/developers/tket}{TKet}, \href{https://www.intel.com/content/www/us/en/developer/tools/quantum-sdk/overview.html}{Intel Quantum SDK}
  \item {\bf Compilation: } Python
  \item {\bf Data set: } Series of blocks of Pauli strings from self-generated UCCSD and QAOA dataset.
  \item {\bf Run-time environment: } Linux and Python.
  \item {\bf Hardware: } Server or PC.
  \item {\bf Metrics: } CNOT gate count, circuit depth, duration, fidelity.
  \item {\bf Output: } Table and figures.
  \item {\bf Experiments: } Compare the metrics on UCCSD dataset.
  \item {\bf How much disk space required (approximately): } 5GB.
  \item {\bf How much time is needed to prepare workflow (approximately): } 10min.
  \item {\bf How much time is needed to complete experiments (approximately): } 24 hours.
  \item {\bf Publicly available: } Yes.
  \item {\bf Code licenses (if publicly available): } MIT License.
  \item {\bf Workflow framework used: } Qiskit, TKet, Intel Quantum SDK.
  \item {\bf Archived (provide DOI): }10.5281/zenodo.10895710

\end{itemize}
}

\subsection{Description}

\subsubsection{How to access} We provide two ways to access the source code. You can download it from the Zenodo link \href{https://zenodo.org/records/10895710}{https://zenodo.org/records/10895710} with DOI 10.5281/zenodo.10895710. You can download the zip file and then decompress it.

You can also run it in docker with the command `docker pull abclzr/vqe\_tetris:latest'.

\subsubsection{Hardware dependencies}

You need a regular server or PC with an intel or AMD CPU.

\subsubsection{Software dependencies} Python, and some python packages: Qiskit, Pytket, pandas, matplotlib.

\subsubsection{Data sets} We provide our self-generated dataset in `core/benchmark/data/'. You don't have to generate the dataset by yourself. But if you want to, we have provided the script in `core/gen\_benchmark.py'.
\subsection{Installation}
\subsubsection{}
If you choose to download the source code from the Zenodo link, you need to run these commands to install Qiskit, Pytket, pandas and matplotlib.
\begin{lstlisting}
pip install qiskit==0.43.1
pip install pytket
pip install pandas
pip install matplotlib
\end{lstlisting}
Then unzip the file, cd to the `vqe\_tetris' folder, and setup the environment variable `PYTHONPATH'. When you open a new terminal, you have to setup `PYTHONPATH' again.
\begin{lstlisting}
unzip vqe_tetris.zip
cd vqe_tetris-master/artifact_evaluation
export PYTHONPATH=../core
\end{lstlisting}
\subsubsection{}
If you choose to run the experiment in a docker container, you firstly need to download the docker image.
\begin{lstlisting}
docker pull abclzr/vqe_tetris:latest
\end{lstlisting}
Start a docker container named `test\_tetris':
\begin{lstlisting}
docker run --name test_tetris -i -d \
abclzr/vqe_tetris:latest
\end{lstlisting}
Now the container is built and detached. Run the following command to attach it to the foreground.
\begin{lstlisting}
docker exec -it test_tetris bash
\end{lstlisting}
You'll see you are in the `/app' directory. cd to `artifact\_evaluation':
\begin{lstlisting}
cd artifact_evaluation
\end{lstlisting}
\subsection{Experiment workflow}
Run the experiments to compare PH and Tetris. This may take 12 hours.
\begin{lstlisting}
python3 run_all.py -test_scale=6
\end{lstlisting}
Run the duration calculation. This may take around 12 hours.
\begin{lstlisting}
python3 calculate_duration.py
\end{lstlisting}
Plot the Fig. \ref{fig:more_baselines} and output Table. \ref{tab:comp2Paulihedral}.
\begin{lstlisting}
python3 show.py
\end{lstlisting}
Run fidelity comparisons.
\begin{lstlisting}
python3 test_fidelity.py
\end{lstlisting}
\subsection{Evaluation and expected results}

The results are stored in `artifact\_evaluation/figs'. If you use docker, you can exit from the container and type 
\begin{lstlisting}
docker cp test_tetris:/app/artifact_evaluation .
\end{lstlisting} to copy the results to your current local directory. In `table2.csv', all the data should be within 5\% relative error w.r.t. Table. \ref{tab:comp2Paulihedral}. In `fig14.pdf', Tetris+lookahead should be the lowest bar. In `fidelity\_LiH.pdf' and `fidelity\_CO2.pdf', Tetris should be higher than PH in most cases.







\bibliographystyle{IEEEtranS}
\bibliography{main}

\end{document}